# MarkerScout: A Disease-Agnostic Machine Learning Framework for Biomarker Prediction from Multi-Scale Mechanistic Models

Robert Moore[1,**], Frank Agayie-Ntim[1,**], Lindsey B. Crawford[1], M. Jana Broadhurst [2], David M. Brett-Major [3,4], Prakash Packrisamy[1], Ahmed Abdeen Hamed[1,*], and Tomáš Helikar[1,*]

[1] Department of Biochemistry, University of Nebraska-Lincoln, Lincoln, NE, US,
[2] Department of Pathology, Microbiology, and Immunology, College of Medicine, University of Nebraska Medical Center, Omaha, NE, United States
[3] Department of Epidemiology, College of Public Health, University of Nebraska Medical Center, Omaha, NE, United States,
[4] Division of Infectious Diseases, Department of Medicine, College of Medicine, University of Nebraska Medical Center, Omaha, NE, United States,

*** Equally contributing first co-authors*

*⁎ Correspondence: ahamed4@nebraska.edu, thelikar2@nebraska.edu,*

**Abstract**

Identifying robust biomarkers from high-dimensional biomedical data is a central challenge in translational research especially for prediction of relevant biomarkers. However, current methods using candidate biomarker rankings produced by any single feature-selection or classification method depend on algorithmic choices and rarely reproducible across pipelines. We present a standalone, disease-agnostic machine-learning framework (MarkerScout) that addresses the issue of hard-wired algorithmic workflows by systematically benchmarking 25 (feature-selection × classifier) pipelines under five-fold stratified cross-validation, aggregating per-feature evidence by two independent methods (a weighted-selection consensus score and Robust Rank Aggregation), and characterizing the direction of each candidate using Cohen's d. We demonstrate the framework on three infectious diseases derived from a companion mechanistic immune-simulation platform: SARS-CoV-2, Influenza A Virus, and Plasmodium falciparum, each evaluated across hospitalization and intensive care unit (ICU) cohorts, yielding six cohorts in total. Best-pipeline cross-validated macro F1 ranged from 0.82 (IAV-HOSP) to 0.99 (COV-ICU), and the framework produces tiered, direction-aware biomarker lists per disease and phase. Interleukin-18 (IL-18) reached the strongest tier in both SARS-CoV-2 phases with consistent direction. Benchmarked against three separate and independently collected clinical ICU datasets, MarkerScout's top-ranked features outperformed 94.4% of randomly selected feature sets of equivalent size for SARS-CoV-2, with a weaker but directionally consistent advantage for Influenza A (66.7%) and *Plasmodium falciparum* (60.7%). The framework is deployed as a REST web service that requires no programming expertise and is applicable to any binary clinical classification problem and supports principled, reproducible biomarker prioritization.

## 1 Introduction

**Motivation**: Biomarker discovery from high-dimensional immune data requires both a biologically grounded patient cohort and a principled method for extracting which features predict clinical outcome, and most existing work addresses only one of these. Understanding how the immune system responds to disease requires models that can faithfully capture the dynamic interplay among hundreds of interacting cell types, cytokines, and regulatory signals across diverse pathological contexts. We recently published a large-scale mechanistic model of the human immune system, validated across 38 independent experiments and spanning 11 disease conditions, that produces exactly such cohorts: simulated patient profiles annotated with clearance outcomes across diseases including SARS-CoV-2, influenza A, and malaria (Moore et al., Front. Immunol., 2026) (1) . What remained was the analytical layer to convert those profiles into ranked, reproducible biomarker candidates. MarkerScout fills that gap: a disease-agnostic framework that benchmarks 25 feature-selection and classifier combinations, aggregates evidence across runs, and assigns tiered, direction-aware biomarker designations from any labeled dataset, whether simulation-derived or clinical.

Biomarkers are now a cornerstone of modern translational research, enabling earlier diagnosis, deeper mechanistic insights, and precise treatments for highly varied diseases. They act as measurable indicators of biological states by capturing the molecular, cellular, and physiological processes that drive disease progression and how a patient responds to therapy. Recent studies show how different biological mechanisms, from immune imbalances to metabolic changes, can be identified through molecular signatures that improve clinical decisions. For example, in acute myocardial infarction (AMI), Miao et al., identified 6 genes associated with cuproptosis and ferroptosis, with expression patterns that distinguished patients from healthy controls (2). Similarly, Grywalska et al., found that higher expression of immune-checkpoint markers and pro-inflammatory cytokines are effective biomarkers in pediatric vulvar lichen sclerosus, illustrating how immune pathways can be used for early detection in chronic inflammatory disorders (3). Together, these findings demonstrate that biomarkers do more than just classify a disease; they reveal the immunological, metabolic, and regulatory circuits that drive pathology, making them essential tools for precision medicine.

Machine learning has become a transformative force in biological discovery. It allows researchers to uncover meaningful structures, mechanistic insights, and predictive patterns within large, complex datasets. Early foundational work showed that supervised and unsupervised learning could effectively classify biological states, infer regulatory relationships, and integrate various omics modalities, building the methodological core of today's computational biology (4–6). As data grew more complex, new frameworks emerged to format, curate, and model high-dimensional molecular information, ensuring that machine-learning algorithms could work effectively with genomic, proteomic, structural, and network-level data (7–9). Deep learning further pushed this boundary by enabling representation learning from raw sequences, images, and single-cell profiles, while advances in large-scale ML frameworks provided the necessary computational power (10–12). Beyond simple prediction, machine learning now drives discovery through network-based (13) and knowledge-guided approaches (14,15), which help identify drug indications, reconstruct regulatory circuits, and integrate mechanistic knowledge into predictive models (16–18). Deep generative models take this even further by learning latent biological structures and proposing novel molecular configurations (19). Ultimately, machine learning is more than just a tool; it is an essential lens that reveals the hidden architecture of biological systems and makes complex experimental questions much more approachable.

Machine learning can detect the subtle, multivariate, and context-dependent patterns needed to identify reliable signatures of disease and therapy response. While traditional statistical methods often struggle with the noise and high dimensionality of 'omics data, machine-learning methods can extract stable, interpretable features. For instance, Hedou et al., developed Stabl, a framework that finds reproducible biomarkers by ensuring stability across resampling, showing that clear signals can be recovered even from noisy clinical data (20). Explainable models have also helped bridge the gap between prediction and actual biological mechanisms. Ijaz et al., for example, used interpretable models to find immune-inflammatory biomarkers for drug-resistant epilepsy (21). Integrating multiple omics data types is another powerful approach; Li et al., showed that tri-omics frameworks can identify biomarkers for sepsis, revealing how immune and metabolic systems fail together during critical illness (22). Other studies have applied ML to immune signatures in joint infections, polycystic ovarian syndrome, and immune-checkpoint pathways like B7x, which moved from discovery toward clinical trials (23). These collective advances show that machine-learning-driven discovery offers a unified way to understand disease, group patients effectively, and personalize care across many medical fields. What these approaches share, however, is a purely observational starting point: signatures are recovered from data that has already been collected, so the predictive candidates that emerge are bounded by the panels that happened to be measured, and the mechanisms linking a marker to an outcome must be inferred after the fact, by default missing the predictive power to assess what is still unknown. A complementary strategy is to begin instead from a system in which those mechanisms are explicit and the measurable space can be defined in advance.

Critically, this mechanistic platform spans 11 distinct disease conditions: nine pathogens: Cytomegalovirus (CMV), Epstein-Barr Virus (EBV), Ebola Virus (EBOV), Human Immunodeficiency Virus (HIV), Mycobacterium tuberculosis (MTB), Helminth, Influenza A Virus (IAV), Plasmodium falciparum (PF), and SARS-CoV-2, as well as Type 1 Diabetes and Lung Transplantation. For each disease, the model can be stimulated to generate large-scale simulated patient cohorts: high-dimensional immune-state profiles in which each record encodes the activation levels of immune cells and secretory factors under cleared or not-cleared infection conditions. These simulated cohorts constitute a structured, disease-labeled dataset immediately suited to supervised machine learning.

Generating simulated patient data, however, is only one side of the biomarker discovery challenge. The other is extracting features from those high-dimensional profiles (cytokines, immune cell populations, and activation states) that most reliably discriminate between clinical outcomes. A single feature selection or classification method is insufficient for this task because algorithmic choices strongly influence which features surface, and rankings produced by any one method rarely generalize across pipelines. As Moore et al. noted, the mechanistic model's simulatable, mechanistically labeled structure positions it as a distinctive substrate for integration with machine learning, offering a principled way to support approaches aimed at identifying immune biomarkers or stratifying immune phenotypes. MarkerScout is that realization.

**Machine Learning Background:** The design of MarkerScout reflects three methodological choices that each have an established empirical basis in the biomarker discovery literature: the use of multiple feature-selection methods rather than a single selector, the factorial benchmarking of classifier families, and the application of stratified cross-validation with selection contained within each training fold.

Selecting which features carry discriminatory signals is not a neutral operation; the choice of method determines which biomarkers surface. Comparative analyses across eight selection methods on DNA microarray data found that different techniques produced substantially different gene sets despite operating on identical evidence, and that disagreement between methods was the rule rather than the exception (Dessì et al., 2013) (24). Stability analyses in multi-omics settings reinforce this: when selection is assessed under repeated cross-validation, the agreement between methods varies widely depending on dimensionality, sample size, and regularization, and no single method consistently outperforms the others across data types (Łukaszuk et al., 2024; He and Yu, 2010) (25,26).

A chemometrics-based evaluation of multiple feature selection algorithms on cancer biomarker data found that agreement in feature lists between methods varied widely, and that high consistency between methods did not guarantee high predictive power, with stability depending on sample size, feature dimensionality, and data quality (Lee et al., 2013) (27). These findings motivate ensemble or multi-method approaches: by aggregating evidence across selectors with complementary inductive biases, the features that surface consistently across methods are more likely to reflect stable biological signals rather than algorithmic preference (Dessì et al., 2013; Li et al., 2022) (24,28).

The classifier applied to a selected feature set introduces a second layer of algorithmic dependence. Comparative benchmarks pairing six classification strategies with seven feature-selection methods on high-dimensional biomedical data showed that no single selector-classifier combination dominated across datasets, and that the best-performing pair varied by data structure rather than following a predictable hierarchy (Wosiak and Dziomdziora, 2015) (29). Similarly, a comparative study of feature selectors and classifiers on three cancer microarray datasets found that simple, well-established selectors paired with appropriately chosen classifiers produced competitive performance without requiring computationally demanding wrapper methods, and that the same selector produced different outcomes depending on the classifier it was paired with (Zanella et al., 2022) (30). More recent work using ensemble feature selection with support vector machines reported that five-fold cross-validated F1 scores above 0.94 were achievable on high-dimensional microarray data, but only after careful matching of the selection strategy to the classifier family (Li et al., 2023) (31). These results support a factorial evaluation design in which multiple selector-classifier combinations are benchmarked in parallel and the best-performing configuration is identified post hoc rather than fixed a priori.

Stratified cross-validation is the standard guard against overfitting and selection leakage in high-dimensional biomarker studies, but its implementation details affect what the resulting estimates actually measure. Benchmark comparisons of feature selection methods on multi-omics data used repeated five-fold stratified cross-validation and found that performance estimates were substantially more stable under repetition than under single-run evaluation, particularly when the number of selected features was small relative to the feature space (Li et al., 2022) (28). Studies using leakage-free nested cross-validation, where feature selection is applied entirely within each training fold and never sees the held-out data, report that naive implementations overestimate performance by a meaningful margin, especially when selection is based on the full dataset before splitting (Pourdadashi et al.,) (32). Cross-validation repeated across multiple Monte Carlo iterations has been used to characterize the variance of selection stability itself rather than just mean predictive accuracy, showing that single-run estimates can misrepresent both the performance and the reproducibility of a feature set (Zobolas et al., 2026) (33). These findings informed the five-fold stratified design used in MarkerScout, in which feature selection is applied independently within each training fold and the held-out fold is used exclusively for performance evaluation.

MarkerScout provides the analytical counterpart to the mechanistic modeling platform: a disease-agnostic machine learning framework that systematically benchmarks 25 (feature-selection x classifier) pipelines under stratified cross-validation, aggregates per-feature evidence across all runs using two independent ranking methods namely: a weighted consensus score inspired by Dutkowski et al., 2007) (34) and Robust Rank Aggregation (35), and assigns tiered, direction-aware biomarker designations grounded in Cohen's d (36–38). Together, the two frameworks constitute an end-to-end computational pipeline for immune biomarker discovery: the mechanistic model generates biologically grounded, disease-specific patient cohorts; MarkerScout converts those cohorts into ranked, reproducible biomarker candidates. In the present work, we demonstrate this combined platform on three of the mechanistic model's disease environments (SARS-CoV-2, Influenza A, and Plasmodium falciparum), spanning six clinical cohorts (hospitalization and ICU phases for each disease), illustrating how the integrated pipeline produces principled, disease-agnostic biomarker predictions at scale. The overall workflow is summarized in Figure 1. Across the six cohorts, best-pipeline cross-validated macro F1 ranged from 0.82 (IAV-HOSP) to 0.99 (COV-ICU); the resulting top-ranked features were then benchmarked against three independently collected clinical ICU datasets, and the framework was deployed as a REST web service to make it usable without programming expertise.

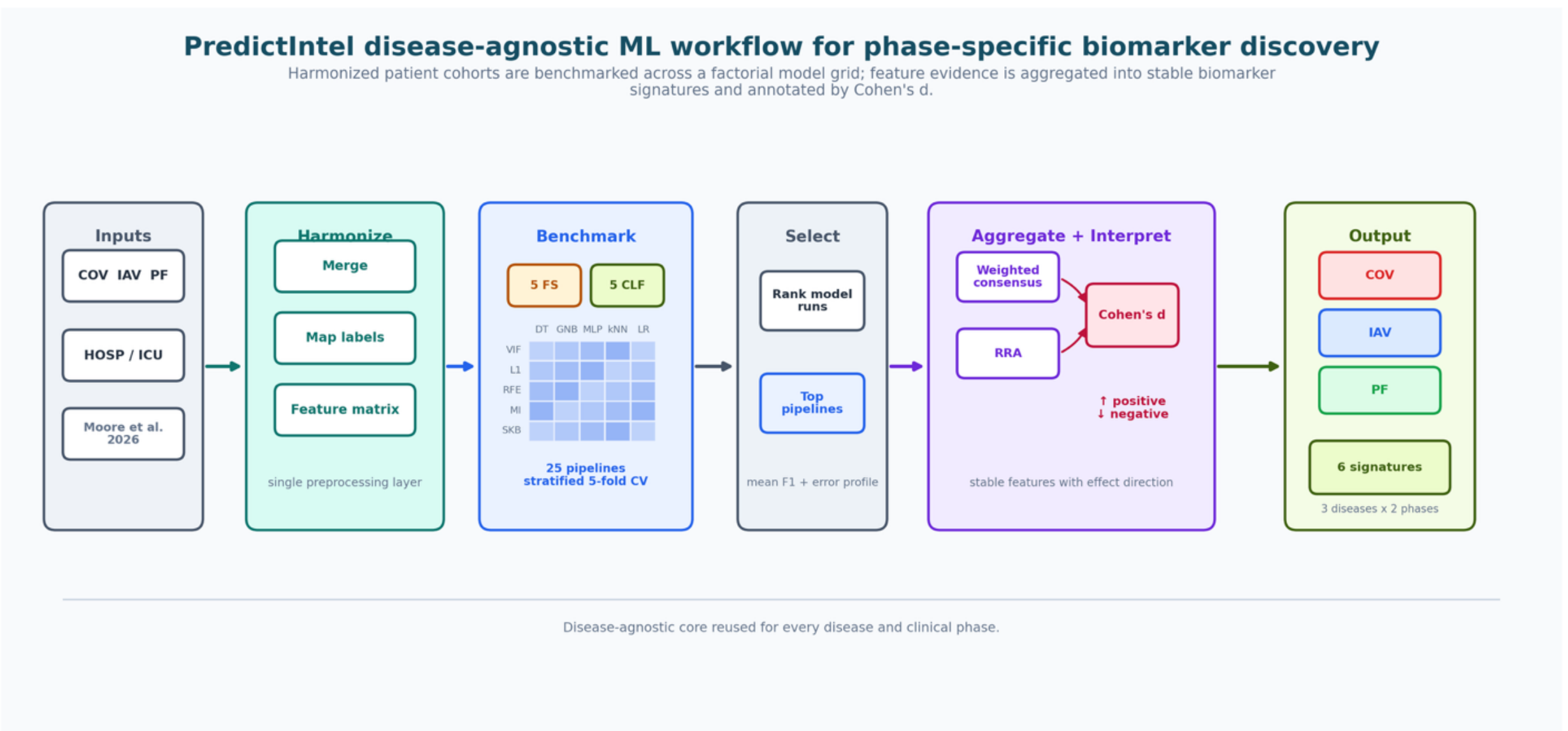

**Figure 1. Adaptive, disease-agnostic machine-learning workflow for phase-specific biomarker discovery.** The pipeline starts with a preprocessing and feature-selection layer to harmonize diverse datasets and outcome labels. It systematically benchmarks 25 (feature-selection x classifier) pipelines using stratified 5-fold cross-validation. Stable biomarker signatures are created by aggregating per-feature evidence using two independent methods: a weighted-selection consensus score and Robust Rank Aggregation. The direction of each candidate feature is characterized using Cohen's d. The combined platform pairs this ML layer with the companion mechanistic immune-simulation model (Moore et al., 2026) to enable end-to-end, disease-agnostic biomarker discovery.

## 2 Materials and Methods

We present a disease-agnostic machine learning framework that predicts the most significant biomarkers that discriminate infection clearance within a given clinical phase (hospitalization or ICU) for any disease, using a labeled patient dataset. The design is adaptive: algorithms are not hardwired. Instead, carefully chosen representatives from each family of feature-selection and classification algorithms are applied to an already-labeled dataset of patient records annotated as cleared or not cleared. Each selector-classifier pair independently predicts the assigned label, and the top-ranked combinations are identified post hoc based on cross-validated performance. Dominant features are then derived from all 25 runs, weighted by cross-validated performance, using two independent aggregation methods, a consensus-based heuristic, and Robust Rank Aggregation, and their directions are characterized using Cohen's d.

Because performance varies across disease and clinical phase, each experiment is treated as independent, and the optimal set of pipelines is determined only after all 25 configurations have been evaluated. Complete pseudocode for each computational step is provided as Algorithms S1–S5 in the Supplementary Algorithms document.

**Framework Inputs.** MarkerScout accepts any tabular dataset in which rows represent individual subjects or simulated records and columns represent biological features, provided two additional inputs accompany it. The first is a binary outcome column in which each record is labeled as either *cleared* or *not_cleared*; this column is the classification target throughout the pipeline. The second is a measurability annotation file: a list of feature names considered clinically or research-grade measurable for the disease and assay context in question. This file acts as a whitelist; only the listed features are carried forward into feature selection and classification, filtering out internal model variables or assay-inaccessible quantities that would be unavailable in a prospective clinical setting. The measurability whitelist is necessarily disease-specific, as the immune components tracked in a SARS-CoV-2 simulation differ from those in an influenza A or *Plasmodium falciparum* simulation, and not all tracked components are measurable with the same assay platforms. The six cohorts analyzed here each carry a disease-specific whitelist reflecting this variation (see Datasets below). A *cleared* record represents a simulated immune trajectory in which the subject resolved infection without progressing to the clinical phase under study; a *not_cleared* record represents a trajectory in which resolution failed and the subject progressed to that phase, with the boundary between the two defined by a percentile threshold on the mechanistic model's Immunoscore calibrated to published epidemiological rates (see Cohorts in Table 1 below).

## 2.1 Datasets

The six cohorts analyzed in this study were constructed from the companion mechanistic immune model (Moore et al., *Front. Immunol.*, 2026) (1) by applying disease-specific percentile thresholds to the model's Immunoscore output. Each threshold was selected to reproduce published hospitalization and ICU admission rates for the corresponding disease, yielding class distributions grounded in real-world epidemiology rather than arbitrary splits. The resulting cleared-to-not-cleared ratios vary substantially across cohorts and directly shape the classification challenge in each experiment; full threshold values and their literature sources are summarized in Table 1.

*Table 1. Immunoscore percentile thresholds and their epidemiological basis. The Immunoscore is an output of the companion mechanistic immune model (Moore et al., Front. Immunol., 2026); percentile cutoffs were applied to this score to assign the binary clearance label for each cohort, with thresholds selected to reproduce published hospitalization and ICU admission rates.*

| Cohort | Percentile Threshold | Epidemiological Rate | Basis |
|---|---|---|---|
| COV-HOSP | 94.3th | 5.7% hospitalized | CDC, pre-vaccination era |
| COV-ICU | 81.5th | 18.4% admitted to ICU | CDC, 2023–2024 |
| IAV-HOSP | 99.05th | 0.95% hospitalized | Real-world IAV hospitalization frequency |
| IAV-ICU | 77th | 23% admitted to ICU | Of hospitalized IAV subset |
| PF-HOSP | 27.5th | 72.5% hospitalized | CDC surveillance data, 2000–2018 |
| PF-ICU | 95th | 5% admitted to ICU | Imported malaria cohorts |

The framework was applied to six cohorts spanning three diseases and two clinical phases each. All cohorts are derived from the companion mechanistic model (Moore et al., Front. Immunol. 17:1732556, 2026; DOI: 10.3389/fimmu.2026.1732556) (1), in which simulated patient trajectories are generated by perturbing immune-component initial conditions across 11 disease contexts. Each record represents one simulated patient profile characterized by a panel of biological features (cell-population activations, cytokine concentrations, and intracellular markers) and is annotated with a binary clearance label (cleared = 0, not_cleared = 1). A feature-measurability annotation indicates which features are considered clinically or research-grade measurable; only these features were retained for downstream analysis.

SARS-CoV-2 (COV). The two COV cohorts are derived from a SARS-CoV-2 immune-response simulation. COV-HOSP comprises 100,000 records described by 108 raw features; after preprocessing, 68 numeric features are retained. COV-ICU comprises 5,673 records with 65 features retained. Class balance in COV-HOSP is approximately 16.6:1 (cleared:not_cleared); in COV-ICU it is approximately 4.5:1.

Influenza A Virus (IAV). IAV-HOSP comprises 100,000 records described by 106 raw features, with 81 retained after preprocessing. IAV-ICU comprises 926 records with 81 features retained. Class balance in IAV-HOSP is approximately 107:1, the most extreme imbalance across all six cohorts; in IAV-ICU it is approximately 3.5:1.

Plasmodium falciparum (PF). PF-HOSP comprises 100,000 records described by 108 raw features, with 75 retained after preprocessing. PF-ICU comprises 71,926 records with 75 features retained. Uniquely among the six cohorts, PF-HOSP exhibits an inverted class balance of approximately 0.4:1 (cleared:not_cleared) (28,074 cleared vs. 71,926 not_cleared), reflecting the high pathogen burden in simulated non-clearing malaria hospitalization. PF-ICU reverts to the cleared-majority pattern at approximately 19.1:1 (cleared:not_cleared), reflecting the high pathogen burden in simulated non-clearing malaria hospitalization. PF-ICU reverts to the cleared-majority pattern at approximately 19:1. Cohort sizes are summarized in Table 2. For each disease, the ICU cohort is derived from the not_cleared subset of that disease's hospitalization cohort, advanced to the intensive-care phase and re-labeled for clearance at that phase. The two phases of a given disease are therefore nested rather than independent, and cross-phase comparisons are interpreted accordingly.

*Table 2. Cohort sizes across six experiments span three diseases and two clinical phases. All cohorts are generated from mechanistic immune-response simulations. Modeled features reflect the feature*

*space after the measurability filter is applied.*

| Cohort | Disease | Phase | Records | Raw Features | Modeled Features After Preprocessing |
|---|---|---|---|---|---|
| COV-ICU | SARS-CoV-2 | Intensive care | 5,673 | 108 | 65 |
| COV-HOSP | SARS-CoV-2 | Hospitalization | 100,000 | 108 | 68 |
| IAV-ICU | Influenza A Virus | Intensive care | 926 | 107 | 81 |
| IAV-HOSP | Influenza A Virus | Hospitalization | 100,000 | 107 | 81 |
| PF-ICU | *P. falciparum* | Intensive care | 71,926 | 109 | 75 |
| PF-HOSP | *P. falciparum* | Hospitalization | 100,000 | 109 | 75 |

### 2.2 Preprocessing

For each cohort, the column set was restricted to the intersection of the measurability whitelist and the non-auxiliary columns. A curated set of columns encoding internal model state (including any feature designated as a resting-state variable) was removed. Target labels were normalized and mapped to a binary encoding (0 = cleared, 1 = not_cleared); records with unmapped target values were explicitly excluded rather than silently dropped. Constant columns, non-numeric columns, and columns with entirely missing values were then removed, and any remaining record containing a missing predictor value was excluded. No imputation was performed. The framework assumes independent and identically distributed observations, consistent with the data's simulated provenance. As summarized in Algorithm S1, each raw disease-phase dataset was transformed into a cross-validation-ready feature matrix and label vector using a standardized preprocessing workflow, including measurability filtering, outcome-label normalization, and removal of incomplete observations.

### 2.3 Feature-Selection Methods

Five feature-selection algorithms were selected to explore the feature space independently, each capturing a different statistical or geometric structure in the data. All methods were configured to retain the top-15 features per fold.

1. SelectKBest: univariate ranking that retains the top-k scoring features by an F-statistic between each feature and the binary target.
2. Mutual Information (MutualInformation): univariate ranking by mutual information between each feature and the target, capturing nonlinear association.
3. RFE with Linear SVM (RFE_LinearSVM): recursive feature elimination under a class-balanced linear support vector machine, applied to standardized training data; the least-informative feature is eliminated at each iteration.
4. L1-Regularized Logistic Regression (L1Selection): sparse logistic regression with class-balanced weights; features with the largest absolute coefficients are retained.
5. Variance Inflation Factor Filtering (VIFFiltering): iteratively removes the feature with the highest variance inflation factor, until at most 15 features remain and all retained features have a VIF below 10.

### 2.4 Classifiers

Five classifiers spanning distinct algorithmic families were evaluated:

1. Logistic Regression: linear baseline with L2 regularization and class-balanced weights.
2. k-Nearest Neighbors (kNN): distance-based voting with five neighbors.
3. Multilayer Perceptron (MLPClassifier): feedforward neural network with a single hidden layer of 100 units and ReLU activation.
4. Gaussian Naive Bayes (GaussianNB): probabilistic classifier operating under a Gaussian feature-likelihood assumption.
5. Decision Tree (DecisionTreeClassifier): hierarchical if-else classifier supporting interpretable decision paths.

The full design yields 25 configurations (feature selection x classifier) per experiment.

### 2.5 Cross-Validation Design

Each of the 25 (feature-selection x classifier) configurations was evaluated under stratified 5-fold cross-validation with a fixed random seed to support reproducibility. Feature selection was fitted exclusively on the training fold and applied to the corresponding held-out validation fold, preventing selection leakage. Predictors were standardized within each fold using the training-fold mean and standard deviation, and classifiers were reinitialized for each split so that no information from earlier folds was carried forward. Algorithm S2 describes the stratified cross-validation framework used to evaluate each feature selector and classifier pair. For each pair, the workflow performs 5-fold stratified cross-validation, records per-fold classification metrics, and tracks feature-selection frequencies across folds.

### 2.6 Per-Fold Metrics and Per-Run Aggregation

For each fold, we computed the two-by-two confusion matrix and derived accuracy, macro-averaged precision, macro-averaged recall, and macro-averaged F1 score. Macro averaging gives equal weight to the cleared and not-cleared classes, which is important given the substantial class imbalances present in the hospitalization cohorts (COV-HOSP roughly 16.6:1; IAV-HOSP roughly 107:1; PF-ICU roughly 19:1) and the inverted balance in PF-HOSP (roughly 2.6:1 not_cleared majority). For each configuration, we report the mean and standard deviation of each metric across the five folds, along with the totals of true positives, true negatives, false positives, and false negatives pooled across all folds.

### 2.7 Cross-Run Aggregation

Three aggregations were performed across the 25 configurations of a single experiment.

**Ranked model summary.** All configurations were ordered by mean F1, with ties broken successively by F1 standard deviation (ascending), macro recall, macro precision, and accuracy.

**Consensus feature ranking.** For each feature, a weighted selection score was computed across all 25 configurations as the sum of the product of its within-run selection frequency and the run's mean F1 score. This ensures that features consistently chosen by higher-performing models contribute more to the final ranking. Algorithm S3 describes the consensus feature-ranking procedure used to assign each feature a weighted selection score. For each feature, the score is computed by summing, across all 25 runs, the product of its within-run selection frequency and the corresponding run-level mean F1 score.

**Robust Rank Aggregation (RRA).** Following Kolde et al. (2012) (21), RRA treats each selector-classifier run as an independent ranked list. For each feature, we compute the beta-distribution-based order-statistic tail probability across its sorted normalized ranks, take the minimum across rank positions as the raw RRA score, and apply Bonferroni correction across contributing runs. Features are ordered by this score (lower = more significant); for reporting, we use the negative base-ten logarithm of the corrected score, with values above 1.3 indicating $p < 0.05$ and above 2.0 indicating $p < 0.01$. Algorithm S4 describes the robust rank aggregation procedure used to combine per-run feature rankings into a consensus feature list. The RRA method evaluates whether each feature's normalized ranks are significantly better than expected by chance.

## 2.8 Discriminatory-Biomarker Tiers

To translate the rankings into a clinically interpretable summary, we partitioned the union of the top 20 features under the Consensus and RRA rankings into three hierarchical tiers. Tier thresholds were chosen a priori and held fixed across all six cohorts to support cross-cohort and cross-disease comparison.

- Tier 1 (Strong): the feature appears in the top 20 of both rankings, has an absolute rank shift of at most 2 positions between rankings, and an RRA score of at least 2.0 (Bonferroni-corrected $p < 0.01$).
- Tier 2 (Moderate): the feature appears in the top-20 of both rankings but fails one Tier-1 criterion (rank shift > 2 or RRA score < 2.0).
- Tier 3 (Exploratory): the feature appears in the top-20 of only one of the two rankings.

Algorithm S5 describes the hierarchical tier-assignment procedure used to classify discriminatory biomarkers from the union of the top-20 Consensus and RRA feature lists. Features are partitioned into three tiers based on rank overlap, rank-shift, and Bonferroni-corrected RRA score thresholds.

## 2.9 Direction of Association

For each tiered feature, Cohen's d was computed on the cleared vs. not_cleared groups using the pooled standard deviation. A positive d indicates a higher mean activation level in the not_cleared group; a negative d indicates a higher mean in the cleared group. Direction labels in figures and tables use the convention "↑ positive / ↓ negative" across all six cohorts, where positive denotes higher activation in the not_cleared group and negative denotes higher activation in the cleared group.

## 2.10 External Clinical Validation

To test whether biomarkers identified from simulated cohorts retain discriminatory value in observed patient data, the tiered features for each disease were compared against three independently published clinical ICU datasets: Lucas et al. for SARS-CoV-2, Cohen et al. for Influenza A , and Dieye et al. for Plasmodium falciparum (39–41). Mechanistic-model feature names were mapped to the analyte names reported in each clinical dataset by direct name matching, and the intersection was retained. The discriminatory performance of the retained MarkerScout features was then evaluated by running the full MarkerScout pipeline (25 feature-selection and classifier configurations under five-fold stratified cross-validation) against the clinical outcome label, and the rank-1 F1 score was compared against a null distribution of 1,000 feature sets of equal size drawn at random without replacement from the remaining measured features in that dataset. The reported statistic is the proportion of the 1,000 random draws for which the MarkerScout feature set outperformed.

### 2.11 Clinical Panel Validation

To assess whether discriminatory signal persists under clinically constrained feature sets, two reduced-panel variants of the COV-HOSP cohort were constructed and run through the full MarkerScout pipeline. The first, the full amenable panel, retains the 13 cytokines from the COV-HOSP feature space that are considered clinically or research-grade measurable and commonly profiled in immunological studies: IFNa, IFNg, IL1B, IL6, IL8, IL10, IL12, IL17, IL18, IL23, IL33, MCP1, and TNFa. The second, the targeted panel, retains the 7 features from that set identified as highest-priority by the full-feature analysis: IFNg, IL6, IL10, IL12, IL17, IL18, and IL33. Both variants were processed with identical preprocessing, cross-validation, and aggregation procedures as the primary COV-HOSP experiment, with no pipeline modifications. The comparison across the three configurations (68 features, 13 features, and 7 features) tests how much discriminatory information is concentrated in the assay-accessible subset of the mechanistic model's feature space.

### 2.11 MarkerScout Web Services APIs

To enable broad accessibility without requiring programming expertise or familiarity with software architecture details, we deployed the MarkerScout framework as a REST web service comprising 13 endpoints. The service supports end-to-end experiment execution (from dataset upload and preprocessing through model training, data visualization, and result retrieval) using standard HTTP API calls. All pipeline endpoints are designed to be asynchronous, anticipating users who submit jobs with large datasets; the caller receives a unique job identifier immediately upon submission and can retrieve results once the background process has completed. Supplementary Table S12 lists the web service endpoints, execution method, path, and required parameters.

The API is organized into three groups: Datasets (dataset ingestion and measurability configuration), Experiments (model training, feature selection, and visualization), and Jobs (asynchronous job submission and status retrieval). Full endpoint specifications are provided in Supplementary Table S12. Figure 5 illustrates the structure of the Experiments endpoint group.

## 3 Results

The framework was evaluated across six cohorts spanning two clinical phases (hospitalization and ICU) for each of three diseases. Across all six cohorts, no single (Selector, Classifier) combination consistently led; five distinct pipelines ranked 1 across the six cohorts, confirming that the framework selected algorithms based on each dataset's signal structure rather than defaulting to a fixed winner. Classification performance varied substantially by disease and phase: SARS-CoV-2 and Plasmodium falciparum cohorts achieved rank-1 mean F1 scores above 0.98, while Influenza A hospitalization produced a rank-1 F1 of 0.82 with a wider top-10 spread, reflecting greater immune complexity and class imbalance at the hospitalization phase. Biomarker Tier-1 set sizes ranged from one feature (IAV-ICU) to seven (IAV-HOSP), a structural variation reflecting differences in how the discriminatory signal is distributed across the feature space. Table 3 summarizes the best pipeline, rank-1 F1, and Tier-1 outputs across all six cohorts. Figure 2 shows the full classification performance matrix across all six cohorts.

*Table 3. Cross–cohort summary of best–performing pipelines and Tier–1 biomarker assignments across all six MarkerScout experiments. Rank–1 F1 is the mean cross–validated F1 score of the top–ranked (Selector × Classifier) pipeline. Tier–1 features satisfy both Consensus and RRA top–20 criteria with rank shift ≤ 2 and $-\log_{10}(p\text{Bonferroni}) \geq 2.0$.*

| Cohort | Best Pipeline | Rank-1 F1 | Tier-1 | Tier-1 Features |
|---|---|---|---|---|
| COV-HOSP | RFE_LinearSVM + LogisticRegression | 0.9916 | 3 | IL10_s, DC_mature, IL18_s |
| COV-ICU | L1Selection + LogisticRegression | 0.9941 | 6 | GMCSF_s, IFNb_s, IL17_s, IL18_s, IL1B_s, IFNa_s |
| IAV-HOSP | L1Selection + MLPClassifier | 0.8216 | 7 | DC_immature, IL1B_s, Mastcell_activated, IL13_s, IFNa_s, IFNb_s, IL4_s |
| IAV-ICU | RFE_LinearSVM + LogisticRegression | 0.9028 | 1 | IL27_s |
| PF-HOSP | RFE_LinearSVM + MLPClassifier | 0.9824 | 4 | IFNg_s, IgE_s, IL3_s, IgM_s |
| PF-ICU | VIFFiltering + MLPClassifier | 0.9879 | 4 | DC_APC, DC_immature, Bcell_Naive, IL23_s |

### 3.1 Framework Behavior Across Cohorts

The extent to which pipeline choice affected performance varied by nearly two orders of magnitude across cohorts. In PF-ICU, the top-10 pipelines spanned 0.002 F1 units, and in COV-HOSP and PF-HOSP 0.008, so that any of the ten configurations would have returned an effectively equivalent classifier. In IAV-HOSP the same span was 0.172 F1 units, an 86-fold difference, where the choice between the first- and tenth-ranked pipeline determined whether macro F1 fell near 0.82 or near 0.65. The two extremes have comparable feature dimensionality (75 and 81 retained features) but differ sharply in class balance (19:1 versus 107:1), indicating that the cost of choosing a pipeline a priori is not fixed but a property of the cohort, not apparent before the benchmark is run.

Five distinct configurations occupied rank 1 across the six cohorts, and the composition of that set is informative. Only three of the five selectors (RFE_LinearSVM, L1Selection, VIFFiltering) and two of the five classifiers (LogisticRegression, MLPClassifier) ever led a cohort; SelectKBest, MutualInformation, kNN, GaussianNB, and DecisionTree did not reach rank 1 in any experiment, although all contributed to the consensus weighting throughout. Embedded and wrapper selectors, therefore, consistently outperformed univariate filters, while no single member of either family led in every case.

The two aggregation methods, applied to the identical set of 25 runs, produced substantially different top-20 lists. The union of the Consensus and RRA top-20 sets comprised 27 features in IAV-HOSP, 29 in PF-HOSP, and 28 in PF-ICU, implying agreement on 13, 11, and 12 of 20 ranked features, respectively. Concordance of 55-65% between two defensible aggregation rules operating on the same evidence is the empirical basis for reporting both rather than either alone, and it is what the tier hierarchy encodes: Tier 1 requires near-identical placement under both rules, whereas Tier 3 captures features that only one rule surfaced.

Behavior under class imbalance was degraded but not broken. At 107:1 in IAV-HOSP the rank-1 pipeline retained macro F1 of 0.8216 with precision 0.9148 against recall 0.7667, a conservative-prediction profile consistent with minority-class under-detection rather than collapse, and cross-fold standard deviation rose to 0.0577. At the milder imbalances of COV-HOSP (16.6:1) and PF-ICU (19:1), rank-1 macro F1 exceeded 0.98 with cross-fold standard deviations of 0.0013 and 0.0017. Cross-fold variability, therefore, tracked separability rather than cohort size.

Tier-1 set size did not track classification performance. IAV-HOSP returned the lowest rank-1 macro F1 in the study (0.8216), along with the largest Tier-1 set (7 features), while IAV-ICU returned higher performance (0.9028) and a single Tier-1 feature. Tier assignment, therefore, measures cross-method rank concordance rather than classification accuracy, and the two are best read as independent axes of evidence: a cohort can be easy to classify while its discriminatory signal is distributed diffusely across many features, or hard to classify while concentrating that signal in a few.

### 3.2 SARS-CoV-2 (COV) Cohorts

**Top-ranked pipelines.** For each SARS-CoV-2 clinical-phase cohort, 25 (Selector × Classifier) pipelines were evaluated under 5-fold stratified cross-validation against the binary clearance outcome.

In COV-HOSP, the top-10 pipelines span a 0.008 F1-unit band (0.9836–0.9916), led by RFE_LinearSVM + LogisticRegression (mean F1 = 0.9916 ± 0.0013, accuracy = 0.9982). COV-ICU is led by L1Selection + LogisticRegression (mean F1 = 0.9941 ± 0.0021, accuracy = 0.9965), spanning a 0.012 F1-unit band. Cross-fold standard deviations never exceeded 0.004 in either cohort, and precision-recall gaps remained ≤0.005, indicating no systematic class bias. Full pipeline rankings are in Supplementary Table S1.

**Error profile.** The rank-1 pipeline simultaneously minimized FP and FN counts in both cohorts: RFE_LinearSVM + LogisticRegression in COV-HOSP (mean FP = 18.6, mean FN = 17.4) and L1Selection + LogisticRegression in COV-ICU (mean FP = 2.0, mean FN = 2.0). FP/FN ratios remained balanced (0.67–1.46) across the top-10 in both cohorts, with error counts scaling proportionally to cohort size. Full error profiles are in Supplementary Table S2.

**Feature scoring and tier assignment.** Features were ranked by Consensus score and RRA (see Methods) and assigned to three tiers.

In COV-HOSP, 3 Tier-1 features were identified: IL10_s, DC_mature, and IL18_s. In COV-ICU, 6 Tier-1 features were identified: GMCSF_s, IFNb_s, IL17_s, IL18_s, IL1B_s, and IFNa_s. IL18_s is the only feature reaching Tier 1 in both COV phases, making it the strongest cross-phase candidate. IL17_s, IFNa_s, IFNb_s, and GMCSF_s appear in Tier 1 at ICU and Tier 2 at HOSP, supporting a phase-modulated role. Complete Tier-2 and Tier-3 assignments are in Supplementary Table S3.

**Direction of association.** To characterize the direction of association independently of the ranking, we computed Cohen's d for each tiered feature on the cleared versus not-cleared groups (a positive d indicates a higher mean in the not-cleared group, while a negative d indicates a higher mean in the cleared group). Among Tier-1 markers, the largest standardized effects in COV-HOSP were IL10_s (d = +2.04), IL18_s (d = -1.16), and DC_mature (d = -1.13); in COV-ICU, the largest effects were IL1B_s (d = +2.68), IFNa_s (d = +2.63), and IFNb_s (d = +2.57).

Twenty features appeared on the tiered lists for both cohorts, and their directions agreed in 19 of 20 cases. IL18_s, IL17_s, GMCSF_s, and IFNg_s trended higher in cleared subjects in both cohorts (negative d), while IFNa_s, IFNb_s, IL10_s, IL12_s, and IL9_s trended higher in not-cleared subjects (positive d). The only direction disagreement was DC_APC, which showed opposite Cohen's d signs between cohorts (HOSP d = +0.04 versus ICU d = -0.58). Since the COV-HOSP magnitude is negligible (|d| < 0.1), this is considered a noise-level disagreement rather than a biological inversion. All five cross-phase candidate markers (IL18_s, IL17_s, IFNa_s, IFNb_s, GMCSF_s) consistently showed the same Cohen's d sign in both phases, reinforcing their cross-phase robustness.

### 3.3 Influenza A Virus (IAV) Cohorts

**Top-ranked pipelines.** In IAV-HOSP, the top-10 pipelines span a 0.172 F1-unit band (0.6497–0.8216), approximately 14–21 times wider than the SARS-CoV-2 cohorts, led by L1Selection + MLPClassifier (mean F1 = 0.8216 ± 0.0172, accuracy = 0.9946). IAV-ICU recovers a tighter range (0.8412–0.9028), led by RFE_LinearSVM + LogisticRegression (mean F1 = 0.9028 ± 0.0244, accuracy = 0.9352). The broader IAV-HOSP spread reflects the greater complexity of influenza immune dynamics at the hospitalization phase. Full pipeline rankings are in Supplementary Table S4.

The elevated cross-fold variability in IAV-HOSP (SD up to 0.0577) reflects severe class imbalance (approximately 107:1 cleared:not_cleared) and a heterogeneous hospitalization-phase immune response. Rank-1 precision (0.9148) substantially exceeded recall (0.7667), indicating conservative prediction of the not_cleared class.

**Error profile.** IAV-HOSP error profiles show pronounced asymmetry: the rank-1 pipeline (L1Selection + MLPClassifier) produced mean FP = 21.4 but mean FN = 86.2, a FP/FN ratio of ~0.25, indicating systematic under-prediction of the not_cleared class. This reflects the greater separability challenge of the influenza hospitalization immune signal relative to SARS-CoV-2. IAV-ICU is less asymmetric: rank-1 produced mean FP = 3.8 and mean FN = 8.2. Full error profiles are in Supplementary Table S5.

**Feature scoring and tier assignment.** In IAV-HOSP, 7 Tier-1 features were identified: DC_immature, IL1B_s, Mastcell_activated, IL13_s, IFNa_s, IFNb_s, and IL4_s. This profile, characterized by innate cellular depletion, type I interferon dysregulation, and Th2-associated cytokines (IL13_s, IL4_s), is consistent with the mixed innate-adaptive signature of influenza hospitalization (42,43) . Complete Tier-2 and Tier-3 assignments are in Supplementary Table S6.

In IAV-ICU, a single Tier-1 feature was identified: IL27_s (consensus rank 7, RRA rank 7, Cohen's d = -0.65), a cytokine with well-established immunoregulatory roles in viral infection. The broad Tier-2 set (10 features) suggests the ICU signal is distributed rather than concentrated. Across phases, IFNa_s and IFNb_s appear in Tier 1 at HOSP and Tier 2 at ICU, making them the strongest cross-phase Influenza A candidates. Complete assignments are in Supplementary Table S7.

**Direction of association.** For IAV-HOSP, the dominant directional pattern is striking in its uniformity: 26 of 27 tiered features trend toward lower activation in subjects who cleared infection (negative d). The single exception is ROS (Tier-2, d = +0.22), which trends higher in not_cleared subjects. The largest effect sizes among Tier-1 markers are IL1B_s (d = -1.32), IL13_s (d = -1.29), and Mastcell_activated (d = -1.21), reflecting substantial suppression of both pro-inflammatory (IL-1beta) and Th2-associated (IL-13) cytokines in those who fail to clear infection. IFNa_s and IFNb_s, while Tier-1, carry smaller effect magnitudes (d = -0.20 and -0.15, respectively), suggesting that type-I interferon levels are informative for classification but not the strongest discriminators of disease severity in IAV-HOSP.

For IAV-ICU, the directional landscape is more heterogeneous. Several features trend higher in not_cleared subjects (DC_pDC, Bcell_Naive, IL1A_s, Th17), while others including IL1B_s, IFNa_s, IFNb_s, and IL27_s trend higher in cleared subjects. This bidirectional pattern is absent in IAV-HOSP and suggests that at the ICU phase, some immune populations are elevated in non-clearing patients (consistent with a hyperactivated or dysregulated immune state), while others are suppressed. The direction reversal of DC_pDC between IAV-HOSP (d = -0.86) and IAV-ICU (d = +0.76) is a notable phase-dependent inversion that warrants biological follow-up.

### 3.4 Plasmodium falciparum (PF) Cohorts

**Top-ranked pipelines.** Both PF cohorts achieved the tightest top-10 F1 spreads among all diseases in this study. PF-HOSP is led by RFE_LinearSVM + MLPClassifier (mean F1 = 0.9824 ± 0.0014, accuracy = 0.9857, top-10 spread = 0.008 F1 units). PF-ICU is led by VIFFiltering + MLPClassifier (mean F1 = 0.9879 ± 0.0017, accuracy = 0.9977, top-10 spread = 0.002 F1 units), with cross-fold SDs ≤0.0024 across the entire top-10. Full pipeline rankings are in Supplementary Table S8.

**Error profile.** PF-HOSP absolute error counts are the largest of any cohort, reflecting the large hospitalization dataset: rank-1 produced mean FP = 101.4 and mean FN = 183.8 (FP/FN ~0.55). PF-ICU rank-1 (VIFFiltering + MLPClassifier) produced mean FP = 19.4 and mean FN = 13.8 (FP/FN ~1.41), a well-balanced error profile. Full error profiles are in Supplementary Table S9.

**Feature scoring and tier assignment.** In PF-HOSP, 4 Tier-1 features were identified: IFNg_s, IgE_s, IL3_s, and IgM_s. The presence of IgE_s and IgM_s in the strong tier reflects the well-established role of immunoglobulin-mediated immunity in malaria, where antibody responses are central to parasite opsonization and clearance. IFNg_s (consensus rank 4, RRA rank 2, d = -2.33) is the dominant pro-inflammatory signal, consistent with its primary effector role in Th1 anti-malarial responses. Complete Tier-2 and Tier-3 assignments are in Supplementary Table S10.

In PF-ICU, 4 Tier-1 features were identified: DC_APC, DC_immature, Bcell_Naive, and IL23_s, a shift from the humoral PF-HOSP profile toward innate antigen-presenting cell populations. DC_immature, Bcell_Naive, and IL1B_s all increased in tier from HOSP to ICU, suggesting these features become more consistently discriminatory as disease severity increases. Complete Tier-2 and Tier-3 assignments are in Supplementary Table S11.

**Direction of association.** In PF-HOSP, 25 of 29 tiered features trend higher in cleared subjects (negative d). The largest Tier-1 effect sizes are IgE_s (d = -2.46), IgM_s (d = -2.37), and IFNg_s (d = -2.33), reflecting strong depression of humoral and gamma-interferon responses in non-clearing patients. RBC_PF (Tier-2, d = +1.87) trends higher in not_cleared subjects, consistent with active parasitemia.

In PF-ICU, 27 of 28 tiered features trend higher in cleared subjects. RBC_PF (Tier-2, d = +2.60) is the sole exception, consistent with active parasitemia at the ICU phase. The largest Tier-1 effects are IL23_s (d = -1.86), DC_APC (d = -1.84), and DC_immature (d = -1.69). Notably, DC_APC reverses direction between cohorts (d = +0.91 at HOSP vs. d = -1.84 at ICU), suggesting a phase-dependent role: early elevation in non-clearing patients at hospitalization followed by depletion at critical illness.

The cross-method agreement between Consensus and RRA rankings across all six cohorts, together with the tier assignment of each feature, is shown in Figure 3. Cohen's d effect sizes for all tiered features across all six cohorts are visualized in Figure 4.

### 3.5 Cross-Phase and Cross-Disease Structure

Because all six cohorts were generated from a single mechanistic platform with a common feature panel and analyzed using an identical pipeline, aggregation, and tier criteria, tier assignments and effect directions are directly comparable across clinical phases and pathogens. The comparisons below are therefore internal to one analysis rather than assembled from independently designed studies, and they describe a structure that cross-phase, cross-pathogen sampling makes visible.

The three diseases did not share a common severity signature. In SARS-CoV-2, failure to clear was associated with elevated pro-inflammatory mediators: the largest Tier-1 effects at the ICU phase were IL1B_s (d = +2.68), IFNa_s (d = +2.63), and IFNb_s (d = +2.57), all higher in non-clearing subjects, with IL10_s (d = +2.04) similarly elevated at hospitalization. Influenza A and Plasmodium falciparum showed the opposite pattern: 26 of 27 tiered features in IAV-HOSP, 25 of 29 in PF-HOSP, and 27 of 28 in PF-ICU were higher in subjects who cleared infection. Under identical analysis, non-clearance therefore presented as a hyperinflammatory state in SARS-CoV-2 and as a hypo-responsive state in influenza and malaria. IL1B_s is the sharpest single instance: the same secreted feature carried d = +2.68 in COV-ICU and d = -1.32 in IAV-HOSP, and also trended higher in cleared subjects at IAV-ICU.

The contrast extended to type I interferon. Despite SARS-CoV-2 and influenza A both being respiratory RNA virus infections represented on the same modeling platform, IFNa_s and IFNb_s reached Tier 1 in both COV-ICU and IAV-HOSP with opposite sign and markedly different magnitude: strongly elevated in non-clearing SARS-CoV-2 subjects (d = +2.63 and +2.57) and weakly elevated in clearing influenza subjects (d = -0.20 and -0.15). In both diseases these features were selected consistently across pipelines, indicating that their classification value does not track effect magnitude alone.

When features reversed direction between clinical phases, the reversals involved dendritic cell populations in every instance. DC_pDC inverted in influenza A, higher in clearing subjects at hospitalization (d = -0.86) and higher in non-clearing subjects at intensive care (d = +0.76). DC_APC inverted in the opposite sense in Plasmodium falciparum, higher in non-clearing subjects at hospitalization (d = +0.91) and higher in clearing subjects at intensive care (d = -1.84). The same feature moved in the same direction in SARS-CoV-2, but at a negligible hospitalization-phase magnitude (d = +0.04 versus d = -0.58 at ICU), which was treated as noise rather than a biological disagreement. Inversions of this kind are not recoverable from single-phase data and would be averaged away by pooling phases.

The composition of the Tier-1 set also shifted with phase in a disease-specific way. In Plasmodium falciparum, hospitalization-phase Tier 1 was anchored by secreted immunoglobulins and IFN-γ (IgE_s, IgM_s, IFNg_s, IL3_s), whereas the ICU-phase Tier 1 comprised cell populations and IL23_s (DC_APC, DC_immature, Bcell_Naive, IL23_s) with no membership in common, a shift from secreted humoral effectors toward antigen-presenting and naive lymphocyte compartments. The tier hierarchy itself moved differently in each disease: SARS-CoV-2 expanded from three Tier-1 features at hospitalization to six at intensive care, with IL17_s, IFNa_s, IFNb_s, and GMCSF_s promoted from Tier 2; influenza A contracted from seven to one, with the residual ICU signal distributed across a ten-feature Tier 2; and Plasmodium falciparum held at four in both phases while replacing its entire membership.

Against this variability, a subset of features was stable. In SARS-CoV-2, 20 features appeared on the tiered lists of both phases, and their directions agreed in 19 of 20 cases; IL18_s, IL17_s, IFNa_s, IFNb_s, and GMCSF_s each held a consistent sign across phases, and IL18_s was the only feature to reach Tier 1 at both. RBC_PF provides an internal check on the platform: as the feature encoding infected erythrocytes, it was higher in non-clearing subjects at both Plasmodium falciparum phases and increased in magnitude with severity (d = +1.87 at hospitalization, d = +2.60 at intensive care), the direction and ordering expected of a parasite-burden measure, and it was recovered by a fully automated pipeline without any modification to the framework's feature selection, classification, or aggregation parameters; the same pipeline that processed SARS-CoV-2 and Influenza A cohorts identified RBC_PF as a top feature in *Plasmodium falciparum* purely from the data.

### 3.6 Validation Against External Clinical Datasets

To assess whether the mechanistic model predictions translated to real patient data, the top features predicted by MarkerScout for each disease were compared against three external clinical datasets collected from ICU patients (see Methods). MarkerScout's predicted features outperformed 944 of 1,000 randomly sampled feature sets for SARS-CoV-2, 667 of 1,000 for Influenza A, and 607 of 1,000 for Plasmodium falciparum. The SARS-CoV-2 result corresponds to an empirical p of 0.056 and represents clear enrichment over chance. The Influenza A and Plasmodium falciparum results are directionally consistent but fall within the range expected under the null hypothesis and are therefore suggestive rather than confirmatory; the most likely contributor is the small overlap between the mechanistic feature space and the panels measured in those two studies, in which only 6 and 2 top-ranked features, respectively, had clinical counterparts, against 15 for SARS-CoV-2 (Table 3). Figure 6 summarizes the composition of feature overlap and benchmarking performance across all three diseases, and Table 3 reports the corresponding counts.

### 3.7 Clinical Panel Validation Results

Restricting the COV-HOSP feature set to the 13-cytokine full amenable panel reduced rank-1 F1 from 0.9916 (SD = 0.0013) to 0.8292 (SD = 0.0883), with L1Selection paired with MLPClassifier emerging as the top pipeline in place of RFE_LinearSVM with Logistic Regression. Further compression to the 7-cytokine targeted panel reduced F1 to 0.7213 (SD = 0.1943), with L1Selection and Logistic Regression leading. The signal degraded with panel size but remained substantially above chance at both configurations. The shift in optimal pipeline across the three experiments is consistent with the framework's adaptive design: as the feature space narrows, the selection geometry changes and different selector-classifier combinations surface. That classification remained viable at 7 features, roughly 10% of the full feature space, indicating that the mechanistic model concentrates meaningful discriminatory information in the assay-accessible nodes it identifies as measurable, rather than distributing it uniformly across all features (Table 4).

*Table 4. COV-HOSP clinical panel validation. Rank-1 F1 and standard deviation across three feature-set configurations: the full 68-feature model, the 13-cytokine full amenable panel, and the 7-*

*cytokine targeted panel. All configurations were run through the identical MarkerScout pipeline under five-fold stratified cross-validation.*

| Configuration | Features | Rank-1 Pipeline | Rank-1 F1 | SD |
|---|---|---|---|---|
| Full model | 68 | RFE_LinearSVM + LogisticRegression | 0.9916 | 0.0013 |
| Full amenable panel | 13 | L1Selection + MLPClassifier | 0.8292 | 0.0883 |
| Targeted panel | 7 | L1Selection + LogisticRegression | 0.7213 | 0.1943 |

## 4 Discussion

Taken together, these comparisons describe a structure that only matched sampling makes visible:The cross-phase and cross-disease comparisons reported above describe a structure that only matched sampling makes visible: pathogen-specific severity signatures, phase-dependent inversions confined to dendritic-cell populations, and disease-specific restructuring of the tier hierarchy between clinical phases.

An independently collected clinical cohort provides external support for the SARS-CoV-2 side of this contrast. In 238 immune-naive, critically ill COVID-19 pneumonia patients enrolled before vaccination and the broad use of immunomodulatory therapy, Brett-Major et al. found all four measured anti-SARS-CoV-2 antibody targets elevated in the most severely affected patients relative to the least, with anti-nucleoprotein IgA the dominant independent predictor (OR 4.3, 95% CI 1.2-20) and a proposed mechanism of IgA-mediated neutrophil activation and NETosis (44). This is the directional logic MarkerScout recovers for the largest SARS-CoV-2 effects, where IL1B_s, IFNa_s, IFNb_s and IL10_s are all elevated in non-clearing subjects: excess immune activity, rather than immune insufficiency, accompanies the worse outcome. The correspondence is sharpened by the contrast with malaria, where the framework places humoral markers (IgE_s, IgM_s) in Tier 1 with negative d, higher in subjects who cleared. Under identical analysis, humoral effectors therefore carry opposite prognostic directions in the two diseases, and the clinical cohort supports the direction assigned to SARS-CoV-2.

Two qualifications apply. The clinical antibodies are antigen-specific (anti-N, anti-RBD) whereas the mechanistic model represents isotype-level secretion, so this is a comparison of direction and mechanism rather than of matched analytes; no feature-overlap benchmarking was performed against that cohort. And the SARS-CoV-2 directional signature is mixed rather than uniform: IL18_s, IL17_s and GMCSF_s trend higher in cleared subjects even as the larger-magnitude markers trend the other way. IL-18 in particular remains discordant with both this cohort's logic and the wider clinical literature, as noted among the limitations.

The hypo-responsive cytokine profile shared by IAV-HOSP and PF-HOSP, characterized by suppressed interferon and inflammatory signaling in non-clearing subjects, may reflect an immunological commonality between the two diseases or a shared feature of the stimulation protocols used to generate those cohorts. Both influenza A and Plasmodium falciparum* involve strong innate immune activation followed by a characteristic suppressive or exhaustion phase, and the mechanistic models encode this trajectory explicitly. The consistency of the hypo-responsive pattern across two immunologically distinct pathogens therefore more likely reflects a genuine biological convergence than a simulation artifact, though sensitivity analyses varying the stimulation protocols would be required to formally exclude the latter.

The mechanistic immune models used here were built exclusively from published human experimental data (Moore et al., 2026) (1) and validated against independent observations(Puniya et al., 2020 and Moore et al., 2026) (1,45), but whether their feature spaces would support robust biomarker discovery in a machine learning context was not established a priori. The results suggest they do. Recovering statistically robust, biologically interpretable biomarker profiles across all six cohorts, each disease-specific in composition and tier structure, without any disease-specific changes to the framework's parameters is difficult to explain unless the models captured the relevant immune dynamics for the clinical phases under study. The framework is agnostic to disease; the biomarker signatures it recovers are not.

The central design choice in MarkerScout was to repurpose classifier F1 scores as feature ranking weights, on the premise that a feature repeatedly selected by high-performing pipelines across algorithmically diverse selectors is more likely to carry real signal than one favored by a single method. The dual-method aggregation, combining the consensus score and Robust Rank Aggregation, was included as a safeguard against the scenario in which one classifier dominated and effectively imposed its geometric preferences on the rankings. The fact that no single (Selector, Classifier) pair led across all six cohorts is reassuring on this point; it also means the framework cannot be reduced to a simpler recommendation about which pipeline to run for a given disease, a design decision we consider a feature rather than a limitation. A formal quantitative assessment confirms this: pairwise Jaccard similarity between the five selectors ranged from 0.13 to 0.39 across cohorts, and pairwise Pearson correlation of mean F1 between the five classifiers ranged from 0.29 to 0.97, with GaussianNB and DecisionTreeClassifier diverging most strongly in the harder IAV and PF cohorts; the cross-phase directional consistency analysis further showed that 95% of shared SARS-CoV-2 features held their Cohen's *d* sign between hospitalization and ICU, compared to 76.9% for *Plasmodium falciparum* and 26.7% for Influenza A, quantifying the phase-dependent instability already described in the results (Supplementary Methods SM1 captures the full analysis**)**.

The scale of the simulated cohorts is an enabling condition of that design rather than a concession. A fully factorial benchmark of 25 pipelines under five-fold cross-validation requires cohorts substantially larger than those available in most clinical biomarker studies: on a cohort of a few dozen patients, per-fold selection frequencies and cross-pipeline rank aggregation are dominated by sampling variance, and a tier structure built on rank concordance would carry little information. Coupling the framework to a mechanistic model that can generate cohorts of 100,000 records is what makes the comparison statistically meaningful, and it permits matched sampling across disease and clinical phase at a scale and balance no single clinical study has achieved. The simulation layer is therefore not a substitute for patient data awaiting replacement, but the component that makes systematic, like-for-like pipeline benchmarking tractable at all.

Performance variation across diseases was informative in itself. The SARS-CoV-2 and *Plasmodium falciparum* cohorts produced tight, high-performing top-10 clusters, indicating that the discriminatory signal in those feature spaces was strong enough for multiple algorithmically diverse pipelines to find it independently; this is the behavior expected of a well-specified mechanistic model whose outputs reflect genuine immune dynamics rather than noise or overfitting to a single algorithmic assumption. Influenza A presented a different challenge: approximately 107:1 class imbalance in IAV-HOSP systematically drove classifiers toward under-predicting the not-cleared minority class, and the wider F1 spread reflected a genuinely harder discrimination problem rather than a framework failure. The tier structure adapted accordingly, producing seven Tier-1 features in IAV-HOSP versus one in IAV-ICU, which is consistent with the framework's intended behavior: stronger cross-pipeline evidence should yield stronger tier assignments, and weaker evidence should not be artificially elevated.

Class imbalance varied substantially across the six cohorts (Table 5). The ICU-phase cohorts were relatively balanced, with cleared-to-not-cleared ratios ranging from 3.5:1 in IAV-ICU to 4.5:1 in COV-ICU. Among the hospitalization cohorts, IAV-HOSP was the most extreme at 107:1 (99,074 cleared vs. 926 not-cleared). PF-HOSP presented a structurally different challenge: not-cleared was the majority class (71,926 vs. 28,074), a cleared:not-cleared ratio of 0.4:1, that nonetheless did not impair pipeline performance, suggesting that the *Plasmodium falciparum* mechanistic model generates a more separable feature space even under majority-class reversal..

*Table 5. Class distribution across all six simulation cohorts. Cleared and not-cleared counts reflect the output of the mechanistic immune model for each disease and severity phase. The imbalance ratio is computed as the majority-class count divided by the minority-class count; a ratio below 1.0 indicates that not-cleared is the majority class.*

| Cohort | Disease | Phase | Cleared | Not-Cleared | Total | Cleared:Not-Cleared |
|---|---|---|---|---|---|---|
| COV-HOSP | SARS-CoV-2 | Hospitalization | 94,327 | 5,673 | 100,000 | 16.6:1 |
| COV-ICU | SARS-CoV-2 | Intensive care | 4,644 | 1,029 | 5,673 | 4.5:1 |
| IAV-HOSP | Influenza A | Hospitalization | 99,074 | 926 | 100,000 | 107.0:1 |
| IAV-ICU | Influenza A | Intensive care | 718 | 208 | 926 | 3.5:1 |
| PF-HOSP | *P. falciparum* | Hospitalization | 28,074 | 71,926 | 100,000 | 0.4:1 |
| PF-ICU | *P. falciparum* | Intensive care | 68,350 | 3,576 | 71,926 | 19.1:1 |

The external clinical validation reported above underscores the same point (Table 6): features identified purely from simulated cohorts retained a measurable discriminatory signal in independently collected patient data, most clearly for SARS-CoV-2.

*Table 6. Validation of MarkerScout top-predicted features against three external clinical ICU datasets. Top features were intersected with each clinical feature set, and their collective discriminatory performance was benchmarked against 1,000 randomly sampled equivalently sized feature sets. MarkerScout's top features outperformed 944/1,000 (94.4 %) random samples for SARS-CoV-2, 667/1,000 (66.7 %) for Influenza A, and 607/1,000 (60.7 %) for P. falciparum.*

| Disease | Clinical Dataset | Total Clinical Features | Overlapping Features | Top Features Overlapped | Random Samples Outperformed |
|---|---|---|---|---|---|
| SARS-CoV-2 | Lucas et al. (Nature, 2020) | 136 | 61 | 15 | 944 / 1,000 (94.4%) |
| Influenza A | Cohen et al. | 37 | 17 | 6 | 667 / 1,000 (66.7%) |
| *P. falciparum* | Dieye et al. | 30 | 13 | 2 | 607 / 1,000 (60.7%) |

The concordance of MarkerScout-predicted features with independently collected clinical data across all three diseases also bears on a question the current study cannot fully resolve: the sensitivity of identified biomarkers to mechanistic model parameterization. The mechanistic models used here reflect a single set of calibrated parameter values; whether varying those parameters within biologically plausible bounds would shift the top-ranked features or leave them largely intact remains to be seen. The clinical validation results provide indirect but meaningful evidence on this point. If the identified biomarkers were artifacts of a particular parameter configuration rather than expressions of genuine immune biology, generalizing from simulation-derived rankings to independently collected patient data would be unlikely. The SARS-CoV-2 result (outperforming 944 of 1,000 random feature sets) is more consistent with the interpretation that the mechanistic model captures biologically grounded immune dynamics than with the alternative that the markers reflect idiosyncratic parameter choices; the weaker Influenza A and Plasmodium falciparum results neither support nor exclude that interpretation. These differences are compounded by substantial variation in cohort class balance across diseases (Table 4.); the 107:1 cleared-to-not-cleared ratio in IAV-HOSP is the most extreme across all six cohorts and likely contributed to the weaker Influenza A result independently of any difference in mechanistic model fidelity. Formal sensitivity analyses of the mechanistic model under parameter perturbation remain a priority for future work and would provide a more direct test of this interpretation.

## 5 Final Remarks and Conclusion

MarkerScout was designed as a disease-agnostic machine learning framework for biomarker prioritization. Its architecture evaluates a fully factorial combination of five feature selectors and five classifiers under stratified cross-validation, using classifier F1 as a feature-ranking weight rather than a model selection criterion alone. No disease-specific parameter changes were made across any of the six experiments; the same framework was applied identically to SARS-CoV-2, Influenza A Virus, and Plasmodium falciparum. Biomarker tier assignments were derived through dual-method ranking aggregation (consensus score and Robust Rank Aggregation), producing a tiered structure in which each level encodes a different degree of cross-pipeline agreement.

Applied to six cohorts spanning the hospitalization and ICU phases across all three diseases, MarkerScout identified tiered biomarker sets for each. For SARS-CoV-2, Tier-1 sets in both phases were dominated by severity-associated cytokines, with IL-18 appearing in Tier-1 at both the hospitalization and ICU levels. For Influenza A, the hospitalization phase recovered innate cellular markers and Th2-associated cytokines in Tier 1, while the ICU phase concentrated signal in a single Tier-1 feature (IL-27), with a broader Tier-2 set reflecting a more distributed immune signature at critical illness. For Plasmodium falciparum, the hospitalization phase Tier-1 was anchored by humoral and Th2-associated markers consistent with antibody-mediated clearance, while the ICU phase shifted to innate antigen-presenting cell populations. This phase-level variation in biomarker structure was recovered from the data without any disease-specific tuning.

The benchmarking of MarkerScout predictions against three external clinical datasets demonstrated that the framework's top-ranked features retained their discriminatory signals in independently collected real-patient data. The SARS-CoV-2 features outperformed 94.4% of randomly sampled feature sets, providing quantitative support for the predictive utility of MarkerScout and underscoring the value of mechanistic immune models as a source of biologically informative biomarker candidates; the Influenza A and Plasmodium falciparum comparisons were directionally consistent but did not reach conventional significance, and warrant validation against larger clinical panels. Future work will extend this framework in two directions: first, to the remaining diseases covered by the companion mechanistic model; and second, to independently collected clinical datasets for conditions not yet represented, applying the same disease-agnostic pipeline to assess generalizability across a broader range of immunological contexts.

## 6 Data and API Availability Statement

The datasets generated and analyzed for this study are available in the MarkerScout GitHub repository. Further inquiries can be directed to the corresponding authors. Users may interact with MarketScout APIs at: ["https://markerscout.gecko-great.ts.net/docs"].

## 7 Author Contributions

Conceptualization, T.H, A.A.H; methodology T.H, F.A.N, and A.A.H.; Investigation, F.A.N, R.M, P.P, L.B.C, A.A.H, and T.H; writing original draft, A.A.H and L.B.C; writing review and editing, A.A.H, L.B.C, R.M, and T.H; funding acquisition, T.H; supervision, A.A.H, T.H.

## 8 Funding

This work was supported by the National Institutes of Health under Grant No. #R35GM119770, and the University of Nebraska-Lincoln Grand Challenges Catalyst Award to Tomáš Helikar.

## 9 Acknowledgments

We are thankful to Skylar Loecker for her valuable discussions during the methodology and investigation phases of this project.

## 10 Conflict of Interest

T.H is a founder and shareholder of Discovery Collective, Inc., and ImmuNovus, Inc.

## 11 Use of AI and AI-Assisted Technologies

Claude (Anthropic) and built-in Gemini in Google Docs were used to reformat and restructure these outputs into manuscript-ready sections. All AI-generated content was reviewed, edited, and validated by the authors, who take full responsibility for the scientific content.

## 12 Limitations

Several limitations bound the scope and interpretation of these findings. All six cohorts are derived from mechanistic model simulations rather than prospectively collected patient data: simulated records are perturbations of a single model structure rather than independent patient observations, the information available to the classifiers is bounded by the 1,450 encoded interactions, and any curation bias in the 449 source publications propagates into the candidate rankings. The near-ceiling F1 scores in the SARS-CoV-2 and *Plasmodium falciparum* cohorts accordingly describe separability within the model's state space and are not estimates of achievable clinical accuracy. Class imbalance varied substantially across cohorts (Table 5), reaching 107:1 in IAV-HOSP, which systematically inflated false negatives even under F1-optimized classifiers. All performance estimates are based on 5-fold cross-validation without an independent held-out partition, so the reported metrics describe internal consistency rather than out-of-sample generalization.

Feature selection was fixed at the top-15 features per fold and tier thresholds were set a priori; neither was tuned, which guards against overfitting but leaves open whether alternative settings would surface different candidates. The external clinical benchmarking was constrained by feature overlap with available clinical panels and reached conventional significance only for SARS-CoV-2; the Influenza A and *Plasmodium falciparum* comparisons were directionally consistent but did not reach significance. Differential performance across diseases may partly reflect unequal mechanistic model fidelity, as the SARS-CoV-2 model drew on a substantially larger experimental evidence base than the Influenza A or *Plasmodium falciparum* models; sensitivity to mechanistic model parameterization, including class ratio choices, was not systematically explored. Performance degrades with panel compression: rank-1 F1 dropped from 0.9916 to 0.8292 at 13 features and to 0.7213 at 7 features in COV-HOSP. Finally, the direction of association reported for some markers is not uniformly concordant with the clinical literature; IL-18 was higher in cleared subjects in the COV cohorts ($d = -1.16$), whereas Lucas et al. 2020 (39) reports IL-18 as generally elevated in severe COVID-19. Model activation states and measured serum concentrations are related but not equivalent quantities, and direction-level validation against clinical data remains outstanding.

## References

1. Moore R, Mayo L, Startsev D, Puniya BL, Poore K, Helikar R, et al. A comprehensive mechanistic multicellular model of the human immune system spanning 11 diseases. Front Immunol. 2026 May 11;17:1732556. doi:10.3389/fimmu.2026.1732556

2. Miao M, Cao S, Tian Y, Liu D, Chen L, Chai Q, et al. Potential diagnostic biomarkers: 6 cuproptosis- and ferroptosis-related genes linking immune infiltration in acute myocardial infarction. Genes Immun. 2023 Aug;24(4):159–70. doi:10.1038/s41435-023-00209-8

3. Grywalska E, Mertowska P, Mertowski S, Zaborek-Łyczba M, Łyczba J, Woźniakowska E, et al. Elevated expression of immune checkpoints and pro-inflammatory cytokines as potential biomarkers in pediatric Vulvar Lichen Sclerosus. Sci Rep. 2026 Feb 2;16(1):4543. doi:10.1038/s41598-025-33630-2

4. Tarca AL, Carey VJ, Chen X wen, Romero R, Drăghici S. Machine Learning and Its Applications to Biology. PLOS Comput Biol. 2007 Jun 29;3(6):e116. doi:10.1371/journal.pcbi.0030116

5. Lee EK. Machine Learning Framework for Classification in Medicine and Biology. In: Van Hoeve WJ, Hooker JN, editors. Integration of AI and OR Techniques in Constraint Programming for Combinatorial Optimization Problems [Internet]. Berlin, Heidelberg: Springer Berlin Heidelberg; 2009 [cited 2026 May 20]. p. 1–7. (Lecture Notes in Computer Science). Available from: http://link.springer.com/10.1007/978-3-642-01929-6_1 doi:10.1007/978-3-642-01929-6_1

6. Ghosh S, Dasgupta R. Machine Learning in Biological Sciences: Updates and Future Prospects [Internet]. Singapore: Springer Nature Singapore; 2022 [cited 2026 May 20]. Available from: https://link.springer.com/10.1007/978-981-16-8881-2 doi:10.1007/978-981-16-8881-2

7. Duran-Frigola M, Fernández-Torras A, Bertoni M, Aloy P. Formatting biological big data for modern machine learning in drug discovery. WIREs Comput Mol Sci. 2019 Nov;9(6):e1408. doi:10.1002/wcms.1408

8. Auslander N, Gussow AB, Koonin EV. Incorporating Machine Learning into Established Bioinformatics Frameworks. Int J Mol Sci. 2021 Jan;22(6):2903. doi:10.3390/ijms22062903

9. Liu X, Zhang W. Bioinformatics in the Age of Big Data: Leveraging Computational Tools for Biological Discoveries. Comput Mol Biol [Internet]. 2024 Aug 25 [cited 2026 May 20];14(0). Available from: https://bioscipublisher.com/index.php/cmb/article/view/3975

10. Ching T, Himmelstein DS, Beaulieu-Jones BK, Kalinin AA, Do BT, Way GP, et al. Opportunities and obstacles for deep learning in biology and medicine. J R Soc Interface. 2018 Apr 4;15(141):20170387. doi:10.1098/rsif.2017.0387

11. Nguyen G, Dlugolinsky S, Bobák M, Tran V, López García Á, Heredia I, et al. Machine Learning and Deep Learning frameworks and libraries for large-scale data mining: a survey. Artif Intell Rev. 2019 Jun 1;52(1):77–124. doi:10.1007/s10462-018-09679-z

12. Goshisht MK. Machine Learning and Deep Learning in Synthetic Biology: Key Architectures, Applications, and Challenges. ACS Omega. 2024 Mar 5;9(9):9921–45. doi:10.1021/acsomega.3c05913

13. Hamed AA, Fandy TE, Tkaczuk KL, Verspoor K, Lee BS. COVID-19 Drug Repurposing: A Network-Based Framework for Exploring Biomedical Literature and Clinical Trials for Possible Treatments. Pharmaceutics. 2022 Mar;14(3):567. doi:10.3390/pharmaceutics14030567

14. Gates LE, Hamed AA. The Anatomy of the SARS-CoV-2 Biomedical Literature: Introducing the CovidX Network Algorithm for Drug Repurposing Recommendation. J Med Internet Res. 2020 Aug 21;22(8):e21169. doi:10.2196/21169

15. Hamed AA, Jonczyk J, Alam MZ, Deelman E, Lee BS. Mining Literature-Based Knowledge Graph for Predicting Combination Therapeutics: A COVID-19 Use Case. In: 2022 IEEE International Conference on Knowledge Graph (ICKG) [Internet]. 2022 [cited 2026 Aug 20]. p. 79–86. Available from: https://ieeexplore.ieee.org/abstract/document/10029987 doi:10.1109/ICKG55886.2022.00018

16. Gilvary C, Elkhader J, Madhukar N, Henchcliffe C, Goncalves MD, Elemento O. A machine learning and network framework to discover new indications for small molecules. PLOS Comput Biol. 2020 Aug 7;16(8):e1008098. doi:10.1371/journal.pcbi.1008098

17. Erbe R, Gore J, Gemmill K, Gaykalova DA, Fertig EJ. The use of machine learning to discover regulatory networks controlling biological systems. Mol Cell. 2022;82(2):260–73.

18. Karpatne A, Kannan R, Kumar V. Knowledge guided machine learning: Accelerating discovery using scientific knowledge and data [Internet]. CRC Press; 2022 [cited 2026 May 20]. Available from: https://books.google.com/books?hl=en&lr=&id=jp0IEQAAQBAJ&oi=fnd&pg=PP1&dq=Karpatne+A,+Kannan+R,+Kumar+V,+editors.+Knowledge+guided+machine+learning:+Accelerating+discovery+using+scientific+knowledge+and+data.+CRC+Press%3B+2022+Aug+15.&ots=aKLQY2rAP6&sig=I_yOw0elUzXFPA9pd6I9YGMZ-cA

19. Lopez R, Gayoso A, Yosef N. Enhancing scientific discoveries in molecular biology with deep generative models. Mol Syst Biol. 2020 Sep 1;16(9):MSB199198. doi:10.15252/msb.20199198

20. Hédou J, Marić I, Bellan G, Einhaus J, Gaudillière DK, Ladant FX, et al. Discovery of sparse, reliable omic biomarkers with Stabl. Nat Biotechnol. 2024 Oct;42(10):1581–93. doi:10.1038/s41587-023-02033-x

21. Ijaz T, Maqsood H, Rehman A, Tahir ul Qamar M, Ashfaq UA. Explainable machine learning identifies immune-inflammatory biomarkers and therapeutic candidates in drug-resistant epilepsy. Sci Rep. 2025 Dec 25;16(1):965. doi:10.1038/s41598-025-30401-x

22. Li Z, Li ZY, Maimaiti Z, Yang F, Fu J, Hao LB, et al. Identification of immune infiltration and immune-related biomarkers of periprosthetic joint infection. Heliyon. 2024 Feb 29;10(4). doi:10.1016/j.heliyon.2024.e26062 PubMed PMID: 38370241.

23. John P, Wei Y, Liu W, Du M, Guan F, Zang X. The B7x Immune Checkpoint Pathway: From Discovery to Clinical Trial. Trends Pharmacol Sci. 2019 Nov;40(11):883–96. doi:10.1016/j.tips.2019.09.008 PubMed PMID: 31677920; PubMed Central PMCID: PMC6907741.

24. Dessì N, Pascariello E, Pes B. A Comparative Analysis of Biomarker Selection Techniques. BioMed Res Int. 2013;2013(1):387673. doi:10.1155/2013/387673

25. Łukaszuk T, Krawczuk J, Żyła K, Kęsik J. Stability of Feature Selection in Multi-Omics Data Analysis. Appl Sci. 2024 Jan;14(23):11103. doi:10.3390/app142311103

26. He Z, Yu W. Stable feature selection for biomarker discovery. Comput Biol Chem. 2010 Aug 1;34(4):215–25. doi:10.1016/j.compbiolchem.2010.07.002

27. Lee HW, Lawton C, Na YJ, Yoon S. Robustness of chemometrics-based feature selection methods in early cancer detection and biomarker discovery. Stat Appl Genet Mol Biol. 2013 Mar 13;12(2):207–23. doi:10.1515/sagmb-2012-0067

28. Li Y, Mansmann U, Du S, Hornung R. Benchmark study of feature selection strategies for multi-omics data. BMC Bioinformatics. 2022 Oct 5;23(1):412. doi:10.1186/s12859-022-04962-x PubMed PMID: 36199022; PubMed Central PMCID: PMC9533501.

29. Wosiak A, Dziomdziora A. Feature Selection and Classification Pairwise Combinations for High-dimensional Tumour Biomedical Datasets. Schedae Informaticae. 2016 Nov 4;2015(Volume 24):53–62.

30. Zanella L, Facco P, Bezzo F, Cimetta E. Feature Selection and Molecular Classification of Cancer Phenotypes: A Comparative Study. Int J Mol Sci. 2022 Jan;23(16):9087. doi:10.3390/ijms23169087

31. Li W, Chi Y, Yu K, Xie W. A two-stage hybrid biomarker selection method based on ensemble filter and binary differential evolution incorporating binary African vultures optimization. BMC Bioinformatics. 2023 Apr 4;24(1):130. doi:10.1186/s12859-023-05247-7

32. Pourdadashi A, Eskandari P, Najafabadi ZG, Tajari A, Heydarzadeh S, Panahi M. Integrated transcriptomic and machine learning analysis identifies female-associated candidate biomarkers in Idiopathic pulmonary arterial hypertension. Comput Biol Chem. 2026 Dec 1;125:109276. doi:10.1016/j.compbiolchem.2026.109276

33. Zobolas J, George AM, López A, Fischer S, Becker M, Aittokallio T. Prognostic biomarker discovery in pancreatic cancer through hybrid ensemble feature selection and multi-omics data.

BioData Min. 2026 Apr 9;19(1):43. doi:10.1186/s13040-026-00546-0 PubMed PMID: 41957754; PubMed Central PMCID: PMC13188360.

34. Dutkowski J, Gambin A. On consensus biomarker selection. BMC Bioinformatics. 2007 May 24;8(5):S5. doi:10.1186/1471-2105-8-S5-S5

35. Kolde R, Laur S, Adler P, Vilo J. Robust rank aggregation for gene list integration and meta-analysis. Bioinformatics. 2012 Feb 1;28(4):573–80. doi:10.1093/bioinformatics/btr709

36. Sullivan GM, Feinn R. Using Effect Size—or Why the P Value Is Not Enough. J Grad Med Educ. 2012 Sep;4(3):279–82. doi:10.4300/JGME-D-12-00156.1 PubMed PMID: 23997866; PubMed Central PMCID: PMC3444174.

37. Diener MJ. Cohen's d. In: The Corsini Encyclopedia of Psychology [Internet]. John Wiley & Sons, Ltd; 2010 [cited 2026 Aug 13]. p. 1–1. Available from: https://onlinelibrary.wiley.com/doi/abs/10.1002/9780470479216.corpsy0200 doi:10.1002/9780470479216.corpsy0200

38. Cohen J. Statistical Power Analysis for the Behavioral Sciences [Internet]. 0 ed. Routledge; 2013 [cited 2026 Aug 13]. Available from: https://www.taylorfrancis.com/books/9781134742707 doi:10.4324/9780203771587

39. Lucas C, Wong P, Klein J, Castro TBR, Silva J, Sundaram M, et al. Longitudinal analyses reveal immunological misfiring in severe COVID-19. Nature. 2020 Aug;584(7821):463–9. doi:10.1038/s41586-020-2588-y

40. Dieye Y, Mbengue B, Dagamajalu S, Fall MM, Loke MF, Nguer CM, et al. Cytokine response during non-cerebral and cerebral malaria: evidence of a failure to control inflammation as a cause of death in African adults. PeerJ. 2016 May 2;4:e1965. doi:10.7717/peerj.1965

41. Cohen L, Fiore-Gartland A, Randolph AG, Panoskaltsis-Mortari A, Wong SS, Ralston J, et al. A Modular Cytokine Analysis Method Reveals Novel Associations With Clinical Phenotypes and Identifies Sets of Co-signaling Cytokines Across Influenza Natural Infection Cohorts and Healthy Controls. Front Immunol. 2019 Jun 18;10. doi:10.3389/fimmu.2019.01338

42. Nguyen THO, Koutsakos M, van de Sandt CE, Crawford JC, Loh L, Sant S, et al. Immune cellular networks underlying recovery from influenza virus infection in acute hospitalized patients. Nat Commun. 2021 May 11;12(1):2691. doi:10.1038/s41467-021-23018-x

43. Lichtner M, Mastroianni CM, Rossi R, Russo G, Belvisi V, Marocco R, et al. Severe and Persistent Depletion of Circulating Plasmacytoid Dendritic Cells in Patients with 2009 Pandemic H1N1 Infection. PLOS ONE. 2011 May 19;6(5):e19872. doi:10.1371/journal.pone.0019872

44. Brett-Major DM, George DS, Morrell ED, Mikacenic C, Carstens JM, Evans LE, et al. Anti-SARS-CoV-2 nucleoprotein IgA is associated with worse disease severity in critically ill COVID-19 pneumonia patients. Front Immunol. 2026 Jul 14;17. doi:10.3389/fimmu.2026.1870217

45. Puniya BL, Moore R, Mohammed A, Amin R, Fleur AL, Helikar T. A comprehensive logic-based model of the human immune system to study the dynamics responses to mono- and coinfections. bioRxiv. 2020 Mar 12;2020.03.11.988238. doi:10.1101/2020.03.11.988238

# 12 Figures

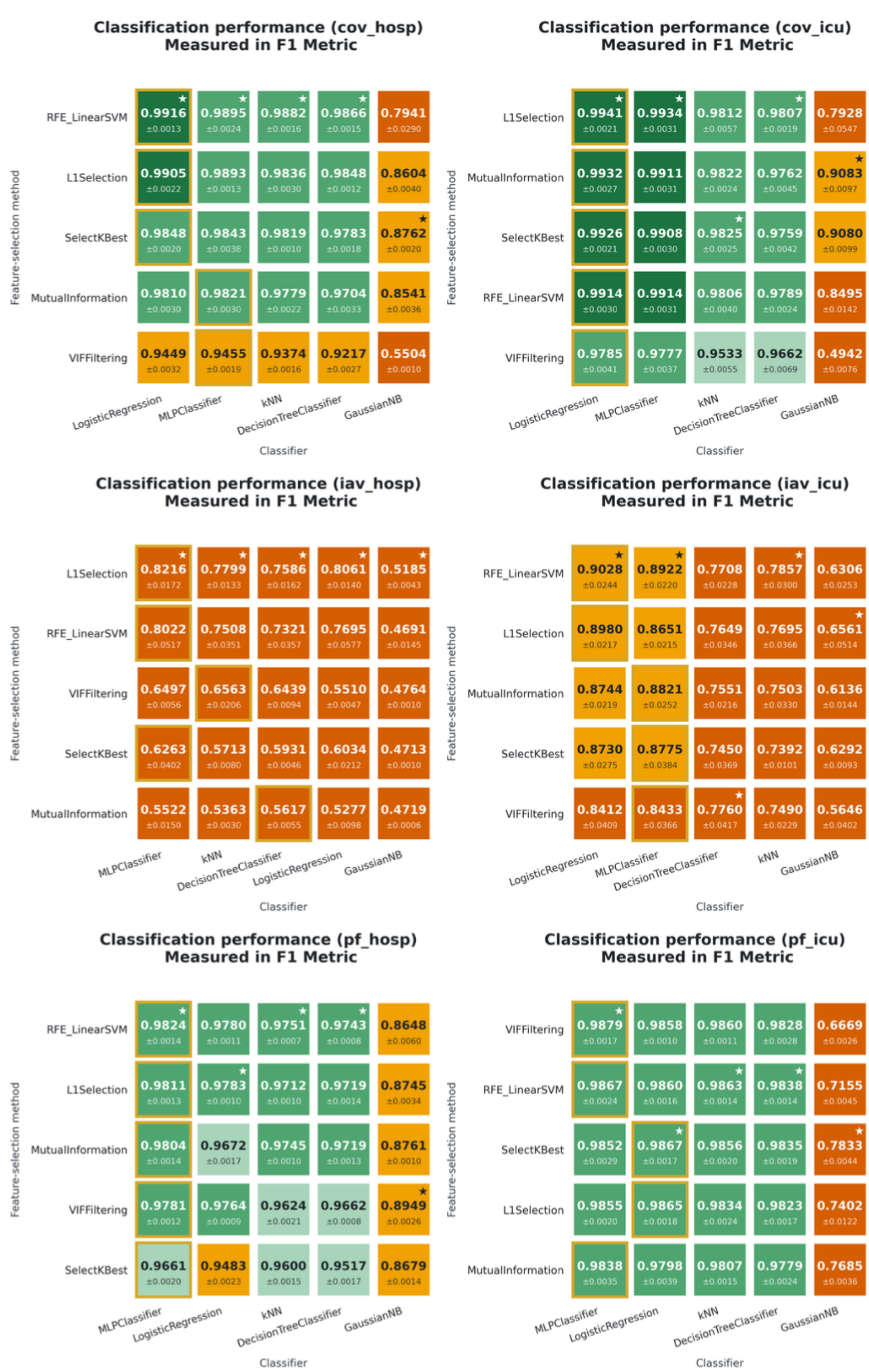

**Figure 2**. Classification performance matrices across diseases and clinical phases. Columns show diseases: SARS-CoV-2 (COV), Influenza A Virus (IAV), and Plasmodium falciparum (PF); rows

show clinical phase: hospitalization (top) and intensive care admission (bottom). Each cell encodes the mean cross-validated F1 score for one feature-selector by classifier combination. SARS-CoV-2 and Plasmodium falciparum cohorts show uniformly high performance across nearly all pipelines, with dense high-F1 clusters and multiple near-equivalent top-performing methods, indicating robust and redundantly recoverable discriminatory signal. In contrast, IAV-HOSP shows the most heterogeneous performance landscape in the study (F1 range 0.65-0.82), consistent with greater immune-feature complexity, whereas IAV-ICU recovers a tighter high-performance band (0.84-0.90). Top-performing cells are visually identifiable as the darkest entries, with RFE_LinearSVM, L1Selection, and LogisticRegression-based combinations recurring among the strongest pipelines across cohorts.

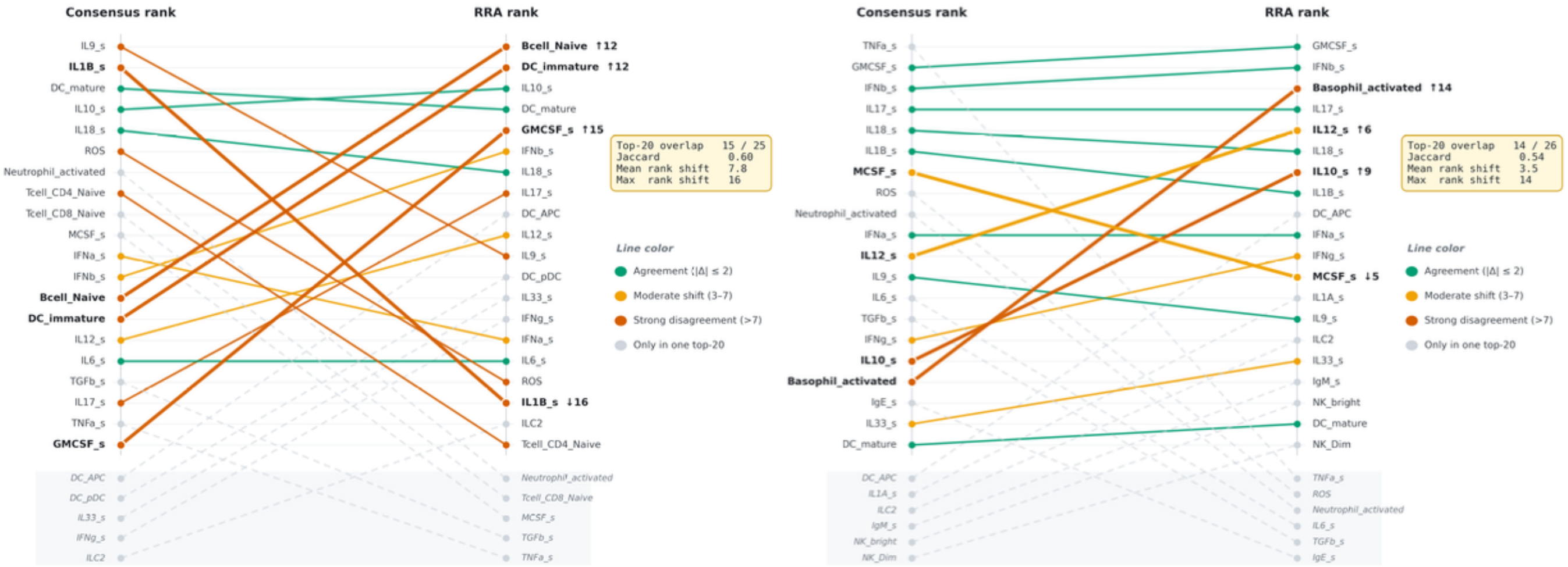

Feature Ranking Agreement (cov_hosp)
Consensus rank
RRA rank
Bcell_Naive ↑12
DC_immature ↑12
GMCSF_s ↑15
IL1B_s ↓16
Top-20 overlap 15 / 25
Jaccard 0.60
Mean rank shift 7.8
Max rank shift 16
Line color
Agreement (|Δ| ≤ 2)
Moderate shift (3–7)
Strong disagreement (>7)
Only in one top-20
Feature Ranking Agreement (cov_icu)
Consensus rank
RRA rank
Basophil_activated ↑14
IL12_s ↑6
IL10_s ↑9
MCSF_s ↓5
Top-20 overlap 14 / 26
Jaccard 0.54
Mean rank shift 3.5
Max rank shift 14
Line color
Agreement (|Δ| ≤ 2)
Moderate shift (3–7)
Strong disagreement (>7)
Only in one top-20

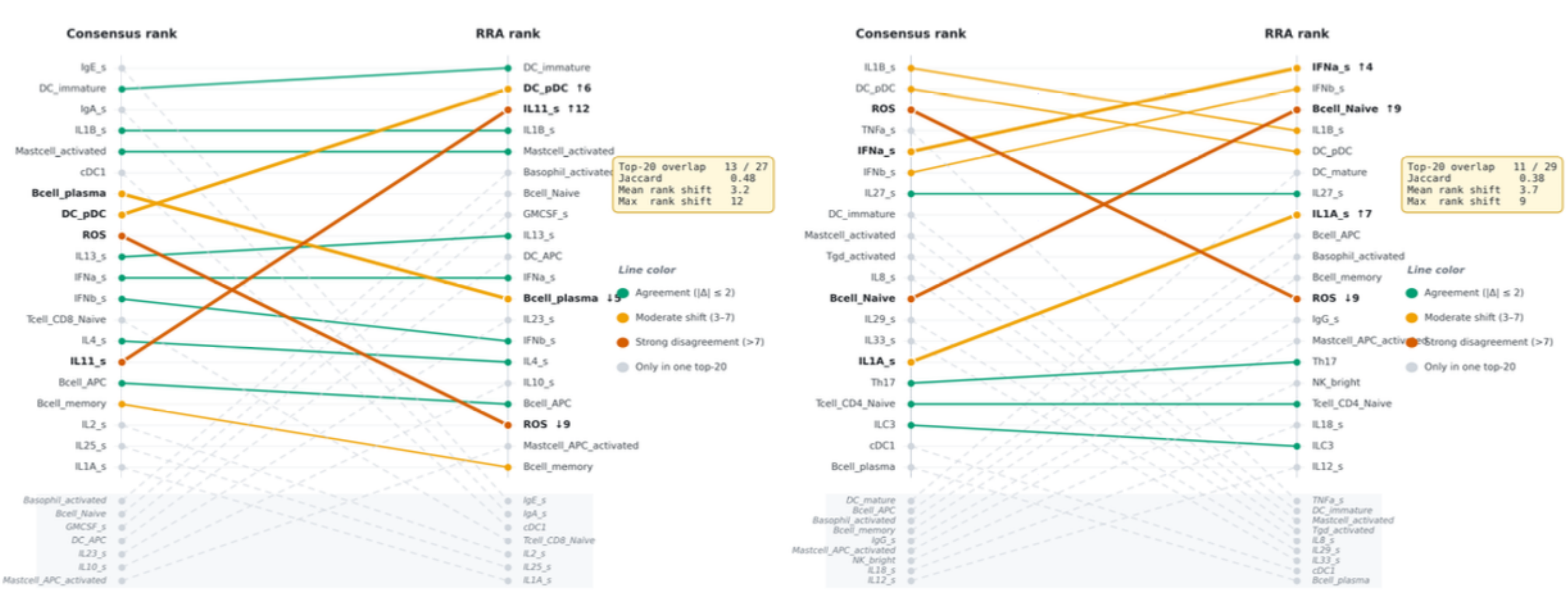

Feature Ranking Agreement (iav_hosp)
Consensus rank
RRA rank
DC_pDC ↑6
IL11_s ↑12
ROS ↓9
Top-20 overlap 13 / 27
Jaccard 0.48
Mean rank shift 3.2
Max rank shift 12
Line color
Agreement (|Δ| ≤ 2)
Moderate shift (3–7)
Strong disagreement (>7)
Only in one top-20
Feature Ranking Agreement (iav_icu)
Consensus rank
RRA rank
IFNa_s ↑4
Bcell_Naive ↑9
IL1A_s ↑7
ROS ↓9
Top-20 overlap 11 / 29
Jaccard 0.38
Mean rank shift 3.7
Max rank shift 9
Line color
Agreement (|Δ| ≤ 2)
Moderate shift (3–7)
Strong disagreement (>7)
Only in one top-20

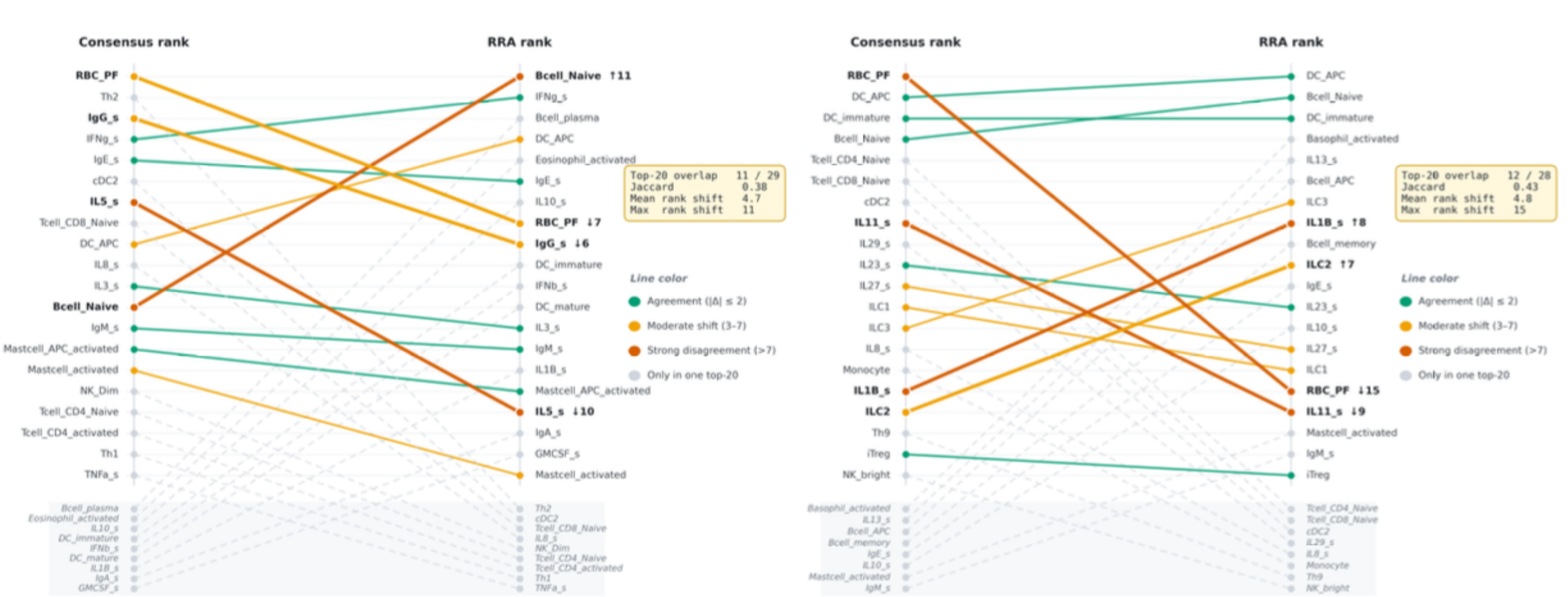

Feature Ranking Agreement (pf_hosp)
Consensus rank
RRA rank
Bcell_Naive ↑11
RBC_PF ↓7
IgG_s ↓6
IL5_s ↓10
Top-20 overlap 11 / 29
Jaccard 0.38
Mean rank shift 4.7
Max rank shift 11
Line color
Agreement (|Δ| ≤ 2)
Moderate shift (3–7)
Strong disagreement (>7)
Only in one top-20
Feature Ranking Agreement (pf_icu)
Consensus rank
RRA rank
IL1B_s ↑8
ILC2 ↑7
RBC_PF ↓15
IL11_s ↓9
Top-20 overlap 12 / 28
Jaccard 0.43
Mean rank shift 4.8
Max rank shift 15
Line color
Agreement (|Δ| ≤ 2)
Moderate shift (3–7)
Strong disagreement (>7)
Only in one top-20

**Figure 3**. Cross-method agreement of top-ranked features across six disease cohorts. Rows: hospitalization (top), intensive care (bottom). Columns: COV (left), IAV (center), PF (right). For each panel, the top-20 features under the Consensus ranking are shown alongside their corresponding ranks under Robust Rank Aggregation (RRA). Bar color indicates tier assignment: green = Tier 1 (strong concordance), orange = Tier 2 (moderate concordance), red = Tier 3 (method-specific).

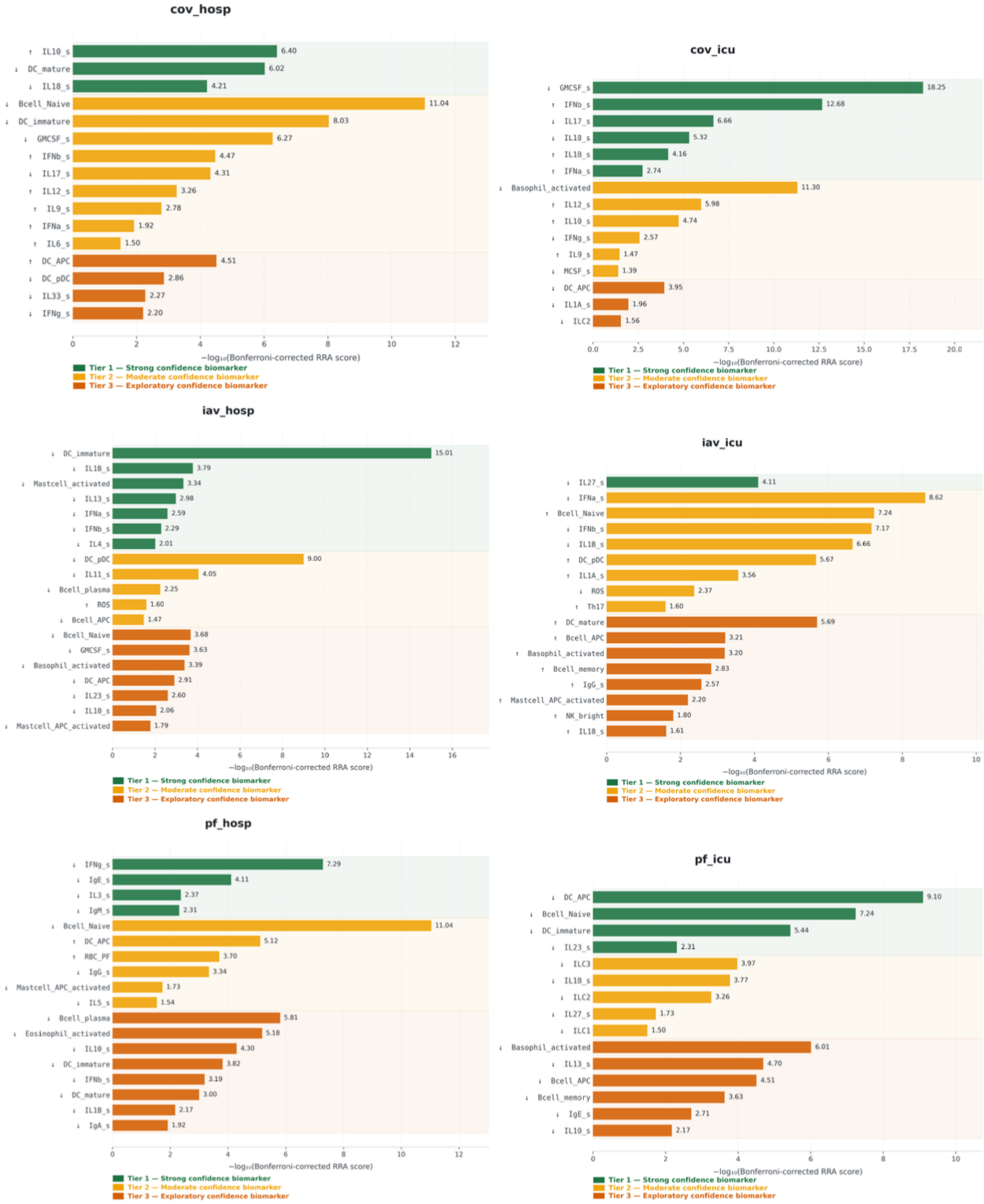

**Figure 4**. Cohen's d effect sizes for tiered biomarkers across six disease cohorts. Rows: hospitalization (top), intensive care (bottom). Columns: COV (left), IAV (center), PF (right). Horizontal bars represent individual tiered features ordered by absolute Cohen's d. Bar color encodes tier: green = Tier 1, orange = Tier 2, red = Tier 3. Bar direction encodes association: right = higher activation in not_cleared subjects; left = higher activation in cleared subjects. IAV-HOSP is dominated by uniformly negative d (lower activation in non-clearing patients) across nearly all tiers, while PF features carry large negative d values at both phases. COV cohorts show a mixed directional signature with the largest effects in IL1B_s, IFNa_s, and IFNb_s (ICU) and IL10_s and IL18_s (HOSP).

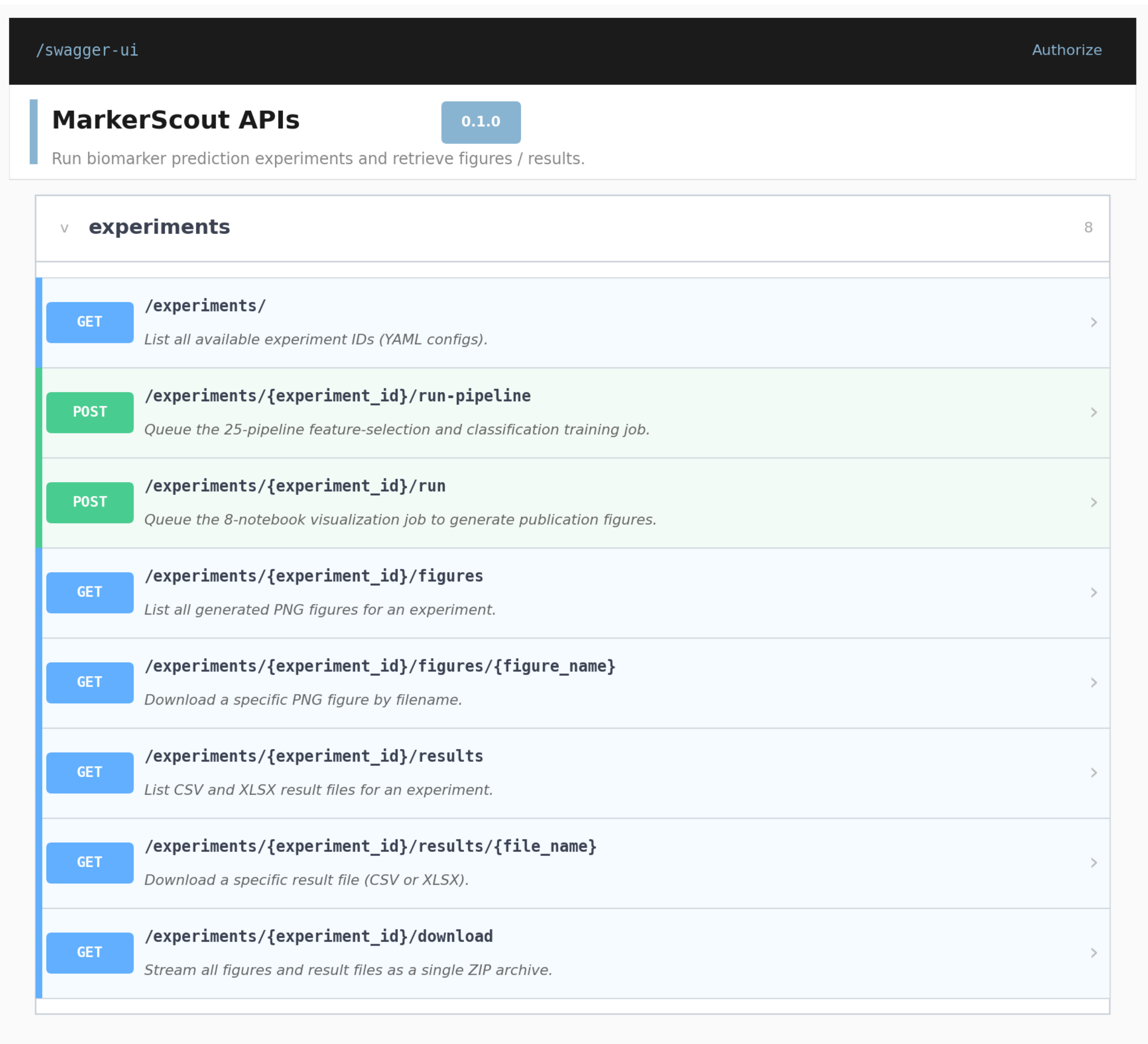

**Figure 5**. MarkerScout REST API experiments endpoint overview. Endpoints organized into experiments where an endpoint consumer may execute the full ML pipeline, retrieve model results, and generate all publication figures.

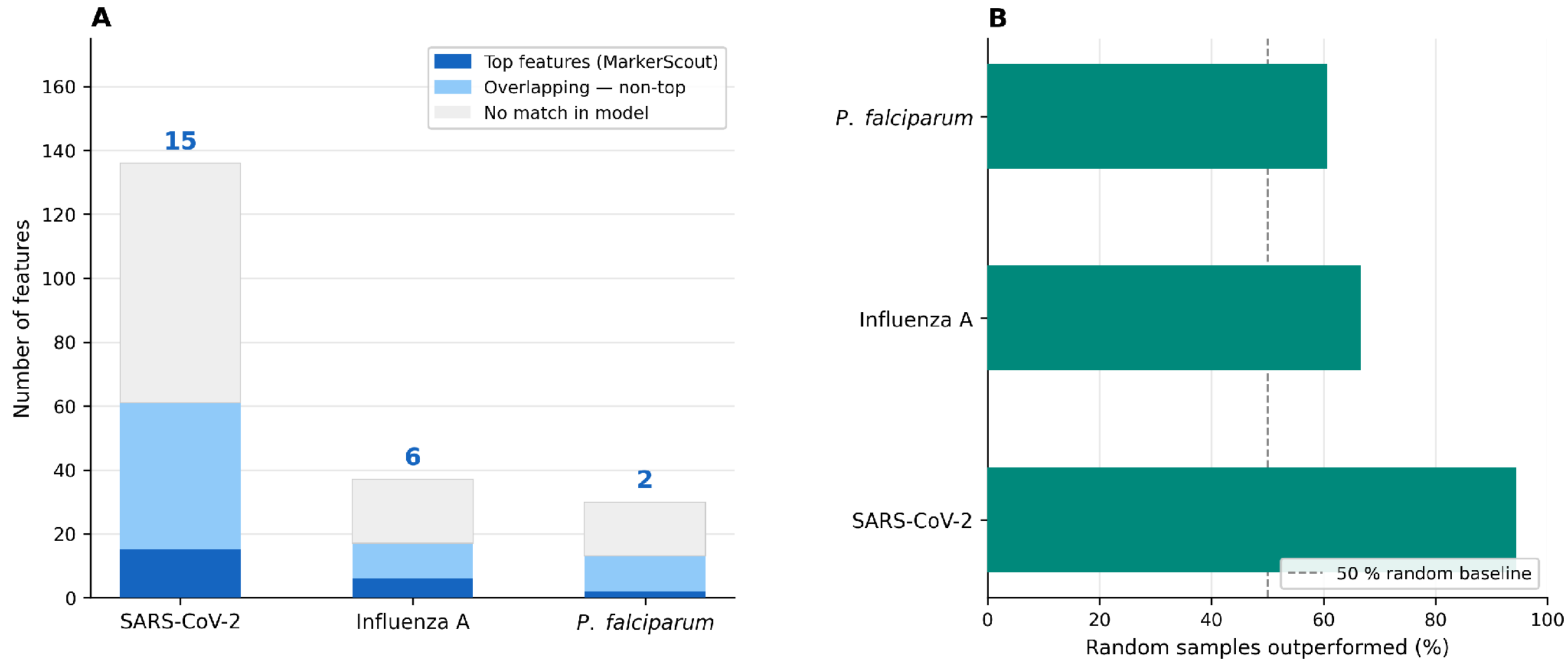

**Figure 6**. Validation of MarkerScout top-predicted features against external clinical ICU datasets. (A) Feature overlap composition: dark blue = MarkerScout top features overlapping with the clinical dataset; light blue = overlapping non-top features; gray = no match in the mechanistic model. Numbers in bars show top-feature counts. (B) Percentage of 1,000 randomly sampled equivalently sized feature sets outperformed by MarkerScout's top features. Dashed line = 50% random baseline.